%% file: main_acm.tex
\documentclass[format=manuscript]{acmart}

\usepackage{booktabs} 
\usepackage[ruled]{algorithm2e} 
\usepackage{subcaption}
\usepackage[printwatermark]{xwatermark}
\usepackage{xcolor}
\usepackage{graphicx}
\usepackage{tikz}
\usepackage{nameref}

\acmJournal{CSUR}
\acmVolume{01}
\acmNumber{01}
\acmArticle{01}
\acmYear{2018}
\acmMonth{01}

\setcopyright{acmlicensed}

\acmDOI{0000001.0000001}

\received{November 2017}
\begin{document}
\title{``Dave\ldots I can assure you \ldots that it's going to be all right \ldots''} 
\titlenote{HAL 9000, \textit{2001 A Space Odyssey}, full quote: ``Just what do you think you're doing, Dave? Dave, I really think I'm entitled to an answer to that question. I know everything hasn't been quite right with me, but I can assure you now, very confidently, that it's going to be all right again.''
 }
\subtitle{A definition, case for, and survey of algorithmic assurances in human-autonomy trust relationships}
\author{Brett W. Israelsen}
\authornote{Corresponding author}
    \orcid{0000-0003-1602-1685}
    \email{brett.israelsen@colorado.edu}
\author{Nisar R. Ahmed}
    \email{nisar.ahmed@colorado.edu}
    \affiliation{%
        \institution{University of Colorado Boulder}
        \city{Boulder}
        \state{CO}
        \country{USA}
    }
    \affiliation{%
        \institution{
        {Cooperative Human-Robot Intelligence  Laboratory (COHRINT)}}
    }
    \affiliation{%
        \institution{
        {Research and Engineering Center for Unmanned Vehicles (RECUV)}}
    }

\begin{abstract}
    People who design, use, and are affected by autonomous artificially intelligent agents want to be able to \emph{trust} such agents -- that is, to know that these agents will perform correctly, to understand the reasoning behind their actions, and to know how to use them appropriately. 
    Many techniques have been devised to assess and influence human trust in artificially intelligent agents. However, these approaches are typically ad hoc, and have not been formally related to each other or to formal trust models. This paper presents a survey of \emph{algorithmic assurances}, i.e. programmed components of agent operation that are expressly designed to calibrate user trust in artificially intelligent agents. 
    Algorithmic assurances are first formally defined and classified from the perspective of formally modeled human-artificially intelligent agent trust relationships. Building on these definitions, a synthesis of research across communities such as machine learning, human-computer interaction, robotics, e-commerce, and others reveals that assurance algorithms naturally fall along a spectrum in terms of their impact on an agent's core functionality, with seven notable classes ranging from integral assurances (which impact an agent's core functionality) to supplemental assurances (which have no direct effect on agent performance). Common approaches within each of these classes are identified and discussed; benefits and drawbacks of different approaches are also investigated. 
\end{abstract}


\thanks{The authors acknowledge the helpful feedback of Michael Mozer and Eric Frew, as well as that of the reviewers. This work was funded by a research gift from Northrop-Grumman Aerospace Systems and by the Center for Unmanned Aircraft Systems (C-UAS), a National Science Foundation Industry/University Cooperative Research Center (I/UCRC) under NSF Award No. CNS-1650468 along with significant contributions from C-UAS industry members.}

\maketitle

\section{Introduction}\label{sec:introduction}
\input{intro.tex}

\section{Motivation and Background} \label{sec:background}
\input{background.tex}

\section{Survey of Algorithmic Assurances} \label{sec:synthesis}
\input{survey.tex}

\section{Future Work} \label{sec:future_work}
\input{future_work.tex}

\vspace{-0.1 in}

\section{Conclusions}\label{sec:conclusions}
\input{conclusions.tex}

\bibliographystyle{ACM-Reference-Format}
\bibliography{References}
\end{document}

%% file: intro.tex
Trust plays a key role in interpersonal relationships. For example, a supervisor asks a subordinate to accomplish a task based on several factors that indicate the subordinate can be trusted to do so. Likewise, when using something like an autonomous vehicle, users must trust it appropriately in order to use it properly. With the rapid advancement of artificially intelligent technology and autonomous systems to do tasks that were previously assumed to be too complicated for machines, there is now much discussion in public \cite{Spectrum2016-jv,DeSteno2014-cq,Wagner2016-ck}, business \cite{Banavar2016-nm, Khosravi2016-ke,Tankard2016-rk}, and academic settings \cite{Foley2017-qj,Castelvecchi2016-mr,Lahijanian2016-nd} on how humans can trust said technology---although, the connection to trust is not always made explicit from a technical standpoint. Those who discuss \emph{how} to trust a specific technology are really referring to the need to identify \emph{indicators of the appropriate level of trust}. 
In other words, it is desirable to \emph{design} capabilities and methods into intelligent technology which help designers, users, and other stakeholders achieve appropriate levels of trust in that technology. These capabilities and methods are collectively referred to as \textit{assurances}. The field of formal Validation and Verification (V\&V) also uses the term assurances to refer to structured evidence that indicates whether or not a system is functioning according to a priori design specifications \cite{Calinescu2017-fh}. These assurances will be referred to here as `hard assurances'. Hard assurances are often not relatable to users of systems or used to adjust levels of trust with a user in real-time, but are used for certification and meeting certain qualifications such as safety. This is in contrast to `soft assurances' that are meant to affect user trust and \emph{trust-related behaviors}. In this paper, `assurances' will refer to soft assurances only.
    
This survey investigates what assurances an \emph{Artificially Intelligent Agent (AIA)} can provide to a human user in order to affect their trust. The colloquial definitions of `appropriate use', `assurance', `AIA', and `trust' should suffice for now to give the reader a general idea of the motivation; more formal definitions will be presented in Section \ref{sec:background}. 
Many researchers from different disciplines will potentially be interested in this work, which includes fields like machine learning, artificial intelligence, robotics and unmanned systems. More broadly, it includes any disciplines that deal in some way with the interface between humans and technology; particularly those who are interested in working with, trusting, interpreting, understanding, and/or regulating AIAs. As such, this paper cuts across multiple disciplines and ties together concepts from several important research topics, such as trustworthy and explainable learning and AI, ethical and transparent autonomy, and safety-/user-aware intelligent systems.

\begin{figure}
    \centering
    \includegraphics[width=0.4\textwidth]{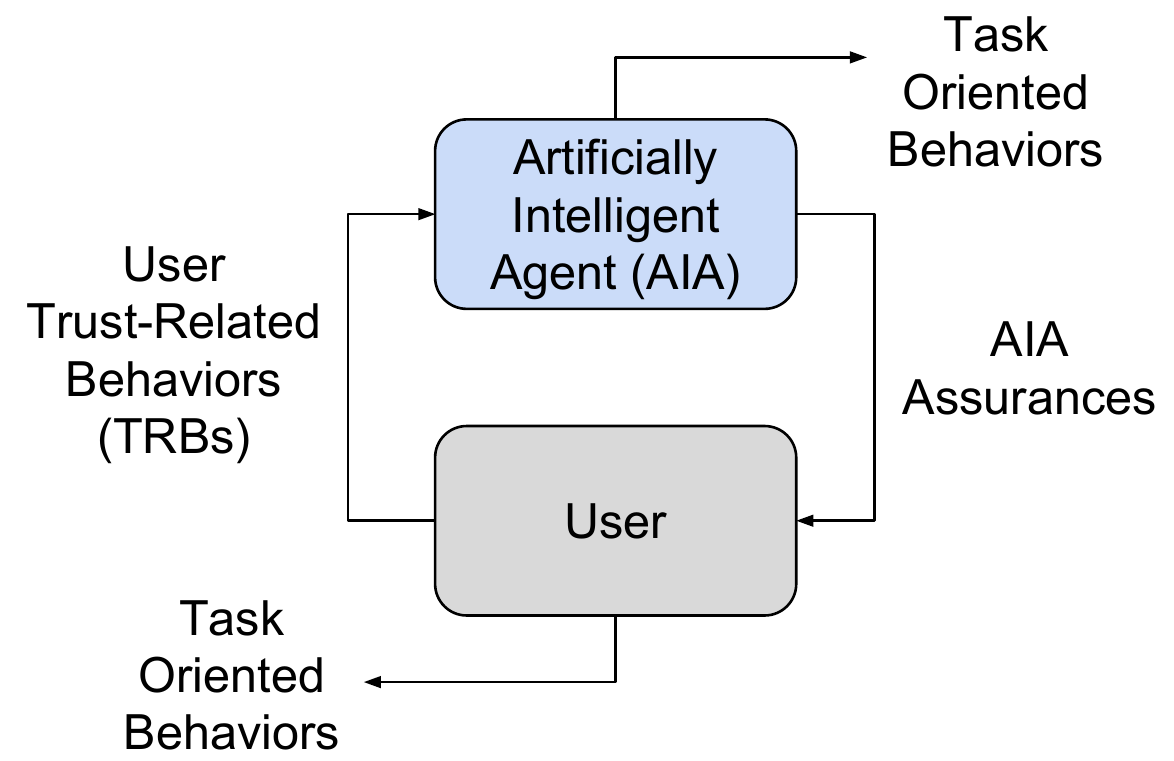}
    \caption{Simple one-way trust relationship between a human user and an AIA: based on a user's level of trust they take certain actions (e.g. give AIA commands), which can lead to AIA actions and/or to assurances which affect user trust. Some `Task Oriented Behaviors' can fall within the trust cycle, but the addition of a separate arrow is meant to encompass any actions are not strictly trust related.}
    \label{fig:SimpleTrust_one_way}
    \vspace{-0.25 in}
\end{figure}

Figure~\ref{fig:SimpleTrust_one_way} is a simple diagram of the trust cycle that exists between a human user and an AIA: user trust is affected by assurances from the AIA, which in turn affects the user's interaction with the AIA (e.g. to trust AIA with responsibilities, or not). To fully understand and appreciate the importance of assurances, one must have a more formal understanding of the components of the trust cycle. This paper provides an overview of the trust cycle elements, and then turns more focused attention to assurances, surveying related research to date. From this survey, properties and classifications of assurances are defined, and considerations for further research are presented. 
Some of the novel contributions of this paper include: creating a detailed description and definition of assurances in general human-AIA relationships (based in research from several, diverse, research communities); making a detailed breakdown of the different components of assurances; and identifying design considerations for implementing assurances at different levels of integration within the AIA. To this end, Section~\ref{sec:background} provides definitions for each of the terms. Section \ref{sec:synthesis} presents a more detailed version of Figure~\ref{fig:SimpleTrust_one_way}. We classify existing work into seven categories of integration within the AIA; detailed definitions, discussion, and examples of assurances from these categories help readers to understand how to apply assurances in their specific applications. Finally, recommendations for future work are discussed in Section~\ref{sec:future_work}, and conclusions are presented in Section~\ref{sec:conclusions}.

%% file: background.tex
The notion of algorithms being used to create assurances for users is \emph{not} new. However, the creation of these assurances (for example in various engineering disciplines, software, science, economics, and others) has historically been done in an ad hoc manner. The need for designed assurances has grown considerably in recent years, as the advanced capabilities of intelligent systems have become more difficult to comprehend and predict \cite{Doshi-Velez2017-xy, Weller2017-zx, Lipton2016-ug, Gunning2017-ih}. Advanced intelligent systems share capabilities with less-advanced counterparts, but generally possess much more delegated responsibility, autonomous functionality, are employed in more uncertain environments, and are operated by a wider demographic of users with different levels of understanding and technical skills. These kinds of technologies are going to be more prolific in number and influence than any other previous technology known to date (consider the number of people already using digital assistants, and content recommendation, as well as the likely impact that autonomous vehicles are likely to have throughout the world). In this atmosphere the practice of designing assurances with little formal understanding is no longer viable; in short: \emph{the existing, informal, approach to assurance design is no longer sufficient due to the new challenges that advanced intelligent systems introduce.}

When researchers discuss concepts like `comprehensible systems', `interpretable learning', `transparent systems', and `explainable AI', they are really interested in making deliberately designed mechanisms to help designers and users appropriately `trust' autonomous and artificially intelligent systems as they perform their tasks. 
For example, many systems are designed to learn from extremely large amounts of data and are expected to regularly perform on never before seen data---yet, it is rarely obvious if such data conforms to assumptions made at design time. 
Other systems are designed to perform tasks that are too `dirty, dull, and dangerous' for humans; the separation of users from these tasks often makes it difficult for them to understand whether these systems are performing as desired. 
The authors, for instance, are interested in the design of unmanned robotic vehicle systems that operate in concert with remote human operators in uncertain dynamic environments. 
Since operators will generally not be computer scientists or roboticists, it is desirable for such systems to behave/communicate in ways that help operators properly use their abilities in scenarios featuring unexpected or incomplete information, time-critical decisions, and risky outcomes~\cite{Hutchins2015-if, Sweet2016-tz}. 
This application is explained in more detail later in relation to Figure~\ref{fig:SimpleTrust_one_way}. 
These issues also have relevance and analogues in other applications of autonomous artificial intelligence, robotics, machine learning and decision making/support systems~\cite{Garcia2015-rs,Otte2013-oo,Sugiyama2013-ci,Amodei2016-xi}, e.g. for scientific data analysis~\cite{Faghmous2014-og}, public policy and medicine~\cite{Wagner2016-ck,Jovanovic2016-gw} and cognitive assistance \cite{Gutfreund2016-xe}.

Some fields have formally and explicitly considered trust between humans and specific forms of intelligent technology, e.g. e-commerce, automation, and human-robot interaction. However, these research efforts have focused largely on developing formal cognitive and psychological models of trust, rather than system behaviors or algorithms that designers can exploit as assurances. 
Other fields that have explored assurance design only provide an informal connection to trust and applications to other disciplines, so it is unknown how effective their developed assurances might be in practice, or what principles ought to be considered for other kinds of autonomous and artificially intelligent systems. 
This paper surveys assurance metrics and methods across relevant application domains, with the goal of
identifying common principles, approaches and questions related to trust-based interaction.
To begin with, definitions for the trust cycle elements in Fig.~\ref{fig:SimpleTrust_one_way} are given to formally ground the concept of assurances. An example application is then provided as a means to compare/contrast technical ideas and implementations of algorithmic assurances throughout the survey in Section~\ref{sec:synthesis}. 


\subsection{Trust Cycle Definitions} \label{sec:trust_definitions}
\input{definitions.tex}


\subsection{Recurring Example Application} \label{sec:mot_example}
    To illustrate the assurances surveyed in the next section, a recurring example application based on the ``VIP escort'' problem~\cite{Humphrey2012-lr} is provided, motivated by the authors' work in unmanned systems. 
    An unmanned ground vehicle (UGV) leads a small convoy through a road network monitored by friendly unattended ground sensors (UGS). The road network also contains a hostile pursuer that the UGV is trying to evade while exiting the network as quickly as possible. 
    The pursuer's location is unknown but can be estimated using intermittent data from the UGS, which only sense portions of the network and can produce false alarms. The UGV's decision space involves selecting a sequence of actions (i.e. go straight, turn left, turn right, go back, stay in place). The UGS data, UGV motion, and pursuer behavior are all stochastic, and the problems of decision making and sensing are strongly coupled: some trajectories through the network allow the UGV to localize the pursuer before heading to the exit (but incur a high time penalty); other trajectories afford rapid exit with high pursuer location uncertainty (increasing the risk of getting caught by the pursuer, which can take multiple paths). 
    A human supervisor monitors the UGV during operation. 
    The supervisor does not have detailed knowledge of the UGV -- but can interrogate its actions, modify its decision making stance (`aggressive' vs. `conservative'), and provide extra information about the pursuer (which is sporadically observed and follows an unknown course). 
    
	\begin{figure}[t]
    	\centering
     	\includegraphics[width=0.4\textwidth]{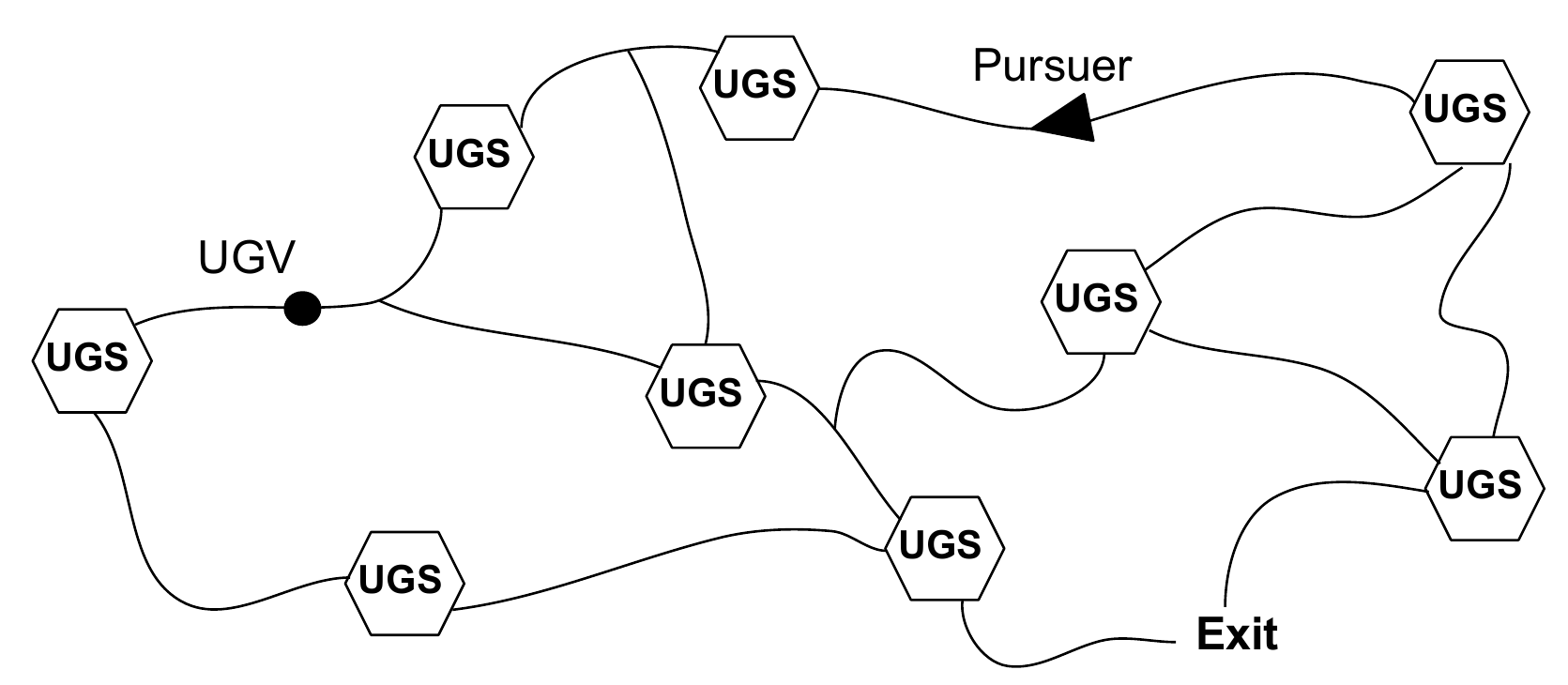}
    	\caption{Application example of unmanned ground vehicle (UGV) in a road network, trying to evade a pursuer, using information from unmanned ground sensors (UGSs), as well as information from a human supervisor.} 
        \label{fig:RoadNet}
        \vspace{-0.2 in}
    \end{figure}

    One way to construct an autonomous UGV path planner is to discretize time and spatial variables to build a partially observable Markov decision process (POMDP) model \cite{Kochenderfer2015-uu} of the navigation task. 
    The ideal POMDP solution is an optimal UGV action selection policy that will, \emph{on average}, maximize some utility function whose optimum value coincides with desirable UGV behaviors (i.e. avoiding the pursuer and exiting quickly). 
    POMDP policies can be calculated by any number of sophisticated approximations that operate on probability distributions for the unknown pursuer state, which in turn can be found via Bayesian sensor fusion \cite{Ahmed2017-ph}. 
    This defines at least two AIA capabilities per Fig. \ref{fig:AIcapabilities}: knowledge representation and planning \footnote{consideration of lower-level UGV state estimation and control also leads to perception and motor control/execution.}. 
    The trust-cycle terms here can then be defined as follows relative to the supervisor (user): \textit{AIA:} the combined POMDP planning and data fusion agent, which must make decisions under uncertainty; 
    \textit{Trust:} the supervisor's willingness to rely on the UGV's planning and data fusion algorithms when the safety of the VIP being escorted is at stake;  
    \textit{TRBs:} supervisor's behaviors that indicate trust (or lack thereof) in the UGV's planner; these include approving/rejecting the planner's actions, or real-time adjustments of the data fusion output based on what the supervisor receives from other intelligence sources; \textit{Assurances:} properties and behaviors of the planning agent that effect the supervisor's trust, e.g. communication of the escape success probability, reports that unexpected UGS data have been registered, or explanations of actions taken.

%% file: definitions.tex
\subsubsection*{Artificially Intelligent Agents} 
\input{aias.tex}
\subsubsection*{User Trust} 
\input{trust.tex}
\subsubsection*{Trust-Related Behaviors} 
\input{trbs.tex}
\subsubsection*{Assurances} \label{sec:assurances}
\input{assurances.tex}

\subsubsection*{Summary}
Each of the elements of Fig.~\ref{fig:SimpleTrust_one_way} has been defined in this Section (\ref{sec:trust_definitions}). Figure~\ref{fig:refined_trust} illustrates how these concepts fit together. In this document algorithmic assurances are surveyed through the lens of their `Level of Integration'; more detailed discussion of the other elements are found in Sec.~\ref{sec:future_work}.

%% file: aias.tex
\begin{figure}[t!]
    	\centering
     	\includegraphics[width=0.55\textwidth]{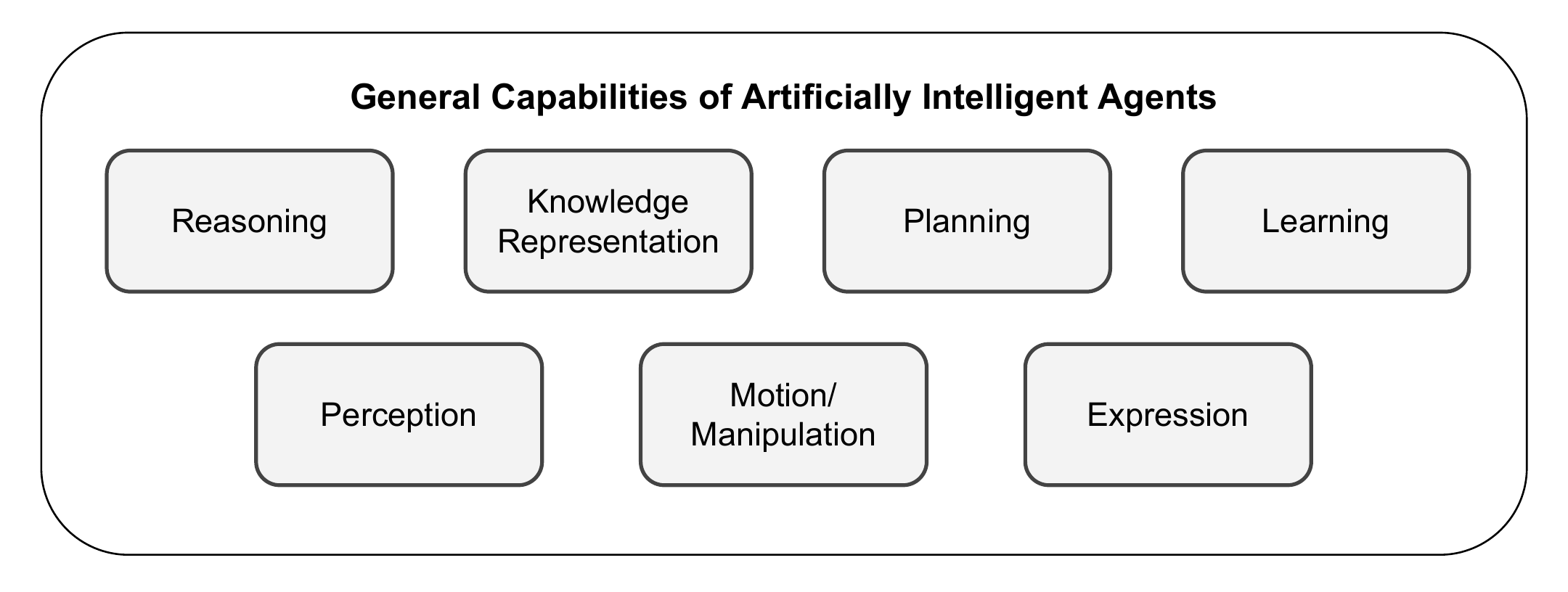}
    	\caption{Set of possible AIA capabilities.}
        \label{fig:AIcapabilities}
        \vspace{-0.2 in}
    \end{figure}

    Herein the term Artificially Intelligent Agent (AIA) will be used in order to encompass a broad range of technologies that can be considered `autonomous'.
    An AIA is defined here as an agent that acts on an internally or externally generated goal, and possesses, to some extent, at least one of the capabilities shown in Fig.~\ref{fig:AIcapabilities} ~\cite{Russell2010-wv,Nilsson2009-rp,Luger2008-vf}. 
    While the term AIA can describe anything from a simple assembly line robot (which only possesses a single capability from Figure~\ref{fig:AIcapabilities}) to the fabled HAL 9000 (who presumably possesses all of the AIA capabilities), this definition underscores the idea that many assurances that exist for one set of (perhaps less capable) AIAs can be adapted and generalized for use in other AIAs.
    In other words, this definition sets a scope for the bodies of research that are likely to have investigated assurances and assurance principles, which can be extended to any `intelligent' computing system. 
    The range of AIA capabilities also helps establish what kinds of assurances might be needed in future systems. 
    For example, assurances for an AIA that only carry out planning tasks will probably differ in design or implementation from assurances for an AIA that only carry out perception tasks. 
    
    It should be noted that an AIA is assumed to operate with a degree of autonomy that is \emph{delegated} by a user. That is, an AIA is self-directed and self-sufficient in its task to the extent that the user's `intent frame' (desired goals, plans, constraints, stipulations and/or value statements) can be met by the AIA, regardless of how it actually accomplishes this. 
    Following \citet{Miller2014-av}, this view of autonomy as a delegation relationship refines the need for `transparent AIAs' by avoiding a contradiction of purpose that stems from an otherwise naive interpretation. From a naive standpoint, one could argue that if AIAs are developed primarily to alleviate the burden of complex reasoning and other undesirable workloads by removing users from the task at hand entirely, then this purpose is undercut by exposure and explanation of sophisticated AIA inner workings to the user. 
    However, if AIAs are subordinates that are delegated tasks by users (who must still act as supervisors), the meaning of `transparency' shifts away from concern over how exactly an AIA accomplishes a task, towards concern over whether or not an AIA can execute the task as per the user's intent frame. 
    This delegation-based view naturally sets up the question of user trust in AIAs. 

%% file: trust.tex
Trust is widely recognized as a critical part of 
intelligent multi-agent system dynamics---from those involving only simple one-on-one interactions \cite{Lewicki2006-hj}, to more complex ones describing markets and governments \cite{Fukuyama1995-un}. 
Because of interest spanning many disciplines, it is difficult (if not impossible) to write a succinct definition of trust that would completely satisfy all interested parties. 
However, following \cite{McKnight2004-vv}, for the purposes of this work trust is defined here as a psychological state in which an agent willingly and securely becomes vulnerable, or depends on, a trustee (e.g., another person, institution, or an AIA), having taken into consideration the characteristics (e.g., benevolence, integrity, competence) of the trustee. 

\begin{figure}[t]
    \includegraphics[width=0.95\textwidth]{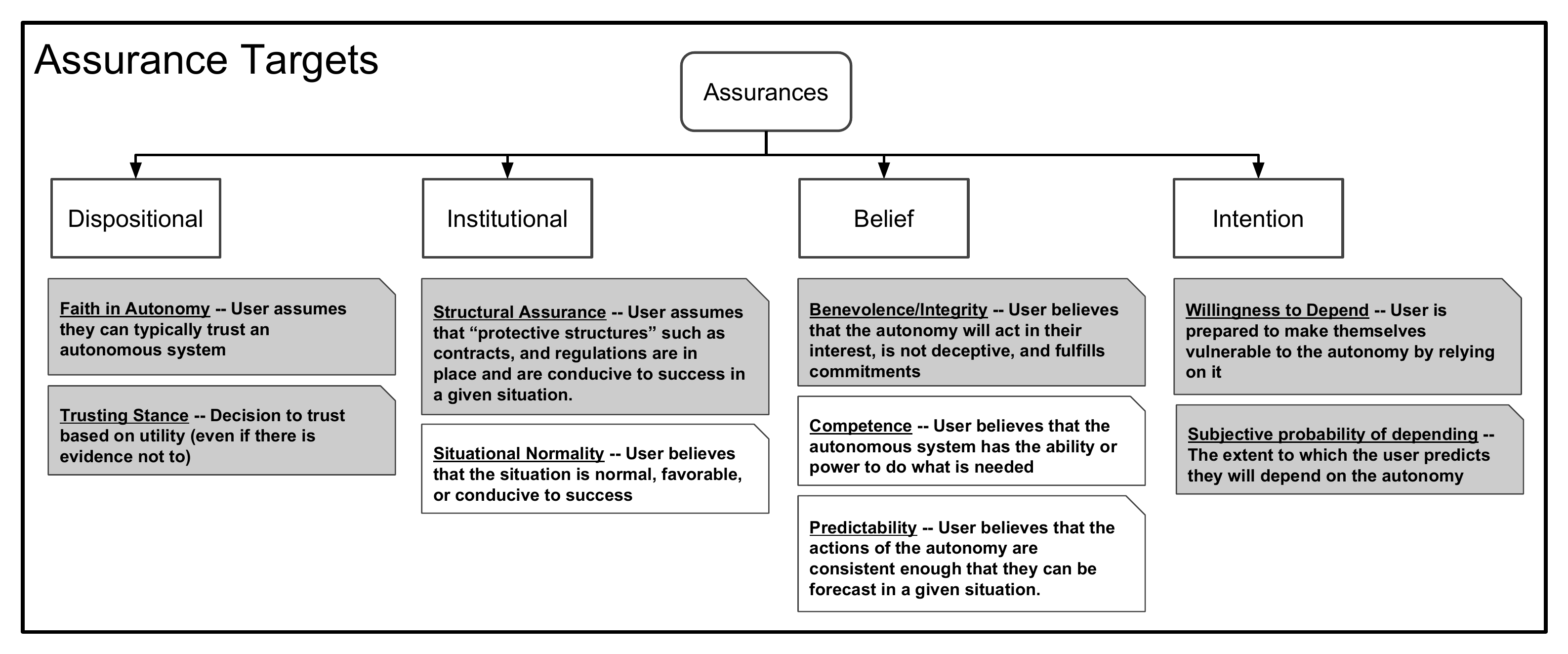}%
    \caption{Notional assurance targets based on the component definitions of the main trust categories. While any of these could be considered targets for assurances, the focus here is only on `Situational Normality', `Competence', and `Predictability'.}
    \label{fig:Assurance_classes}
    \vspace{-0.2 in}
\end{figure}

This raises two important questions. Firstly, since trust is generally understood to exist between people, is it possible for a human to enter into a trusting relationship with an AIA? 
That humans actually do develop trust in autonomous machines has been confirmed several times~\cite{Muir1996-gt,Mcknight2011-gv,Riley1996-qm,Bainbridge2011-pl,Salem2015-md,Desai2012-rc, Freedy2007-sg, Kaniarasu2013-ho, Wang2016-id}. 
\citet{Lacher2014-yc} also notes that people trust AIAs for transportation systems at different levels (i.e. an engineer trusts differently than an operator or passenger). 
Secondly, in designing assurances that affect trust-based user behaviors, is it possible to know what drives those behaviors and thus have some working model of user trust that can be mapped to AIAs? 
\citet{McKnight1998-ty} (and later \cite{McKnight2001-fa}) performed what is arguably the first multi-disciplinary survey and unification of trust literature, which also condensed it into a single typology consisting of four major related components. 
Adapted to AIAs, these are: \textit{Disposition to Trust:} the extent to which one displays a tendency to be willing to depend on AIAs in general across a broad spectrum of situations and persons; \textit{Institution-Based Trust:} the extent to which one believes that regulations are in place that are conducive to situational success in an endeavor; \textit{Trusting Beliefs:} the extent to which one believes that the AIA has one or more characteristics beneficial to oneself; \textit{Trusting Intentions:} the extent to which one is willing to depend on, or intends to depend on, the AIA even though one cannot control its every action. 
Dispositional Trust is generally considered by psychologists, and deals with long-term psychological traits that develop in a person from childhood (e.g. is someone pre-disposed to trusting technology?).  
Institutional Trust 
is generally studied by sociologists, and represents the level to which a person trusts social/commercial structures. 
Finally, Interpersonal Trust (encompassing both `trusting beliefs', and `trusting intentions') deals directly with one-on-one relationships and tends to fluctuate most quickly. 
Each of these trust components has sub-components defined in Figure~\ref{fig:Assurance_classes}, which were identified by compiling many research studies across several disciplines. 
These components are the principal drivers of user trust-related behaviors, and are the general notional targets of AIA assurances.

%% file: trbs.tex
Trust ultimately leads to some kind of meaningful behavior or action which reflects the level of an individual's trust \cite{Lewis1985-pr}. 
These actions are called `trust-related behaviors' (TRBs) \cite{McKnight2001-fa}. 
In the case of a human-AIA relationship per Fig. \ref{fig:SimpleTrust_one_way}, 
some example TRBs could include the kinds of tasks the human user assigns to the AIA, accepting and following through on a plan produced by the AIA, directing that a new plan be made, or switching off autonomous capabilities altogether to teleoperate and perform tasks manually through a physical mechanism that the AIA otherwise controls.  

Trust is not a univariate quantity that can be objectively measured. Rather, it is a multidimensional phenomenon whose `relative magnitudes and directions' must be observed through changes in TRBs, or qualitative self-reports gathered via surveys \cite{Muir1996-gt}. It thus comes as no surprise that TRBs are the more objective method of observation due to the fact that people are not always consistent in their ratings, and may sincerely feel different levels of trust while performing similar TRBs~\cite{Dzindolet2003-ts}. \citet{Parasuraman1997-co} were interested in understanding the use of automation by humans, and defined terms to describe that use. Here it is proposed that, by extension, those terms also apply to the behaviors of humans towards more advanced AIAs. Within this scope the definitions are as follows: \textit{Misuse:} over-reliance on an AIA (which could manifest itself in a user's unrealistically optimistic expectations of performance); \textit{Disuse:} under-utilization of an AIA (e.g. a user turning off the AIA prematurely, or failing to use all of its capabilities); \textit{Abuse:} Inappropriate application of an AIA (where \emph{application} in this case means the choice to deploy an AIA in a certain context).

Following Fig.~\ref{fig:SimpleTrust_one_way}, AIA assurances should ideally be designed to steer the user away from misuse, disuse, or abuse of the AIA, i.e. towards otherwise appropriate TRBs, by properly `calibrating' assurances to suitably influence user trust. This point, to some extent, has been alluded to in \cite{Muir1994-ow,Lillard2015-yg,Lee2004-pv,Hutchins2015-if}. Other researchers who propose `calibration' (or related concepts) suggest calibrating \emph{trust} as opposed to TRBs. \citet{Dzindolet2003-ts} found that providing system performance feedback tended to increase users' \textit{self-reported trust}, even though resulting TRBs did not reflect self-reported trust levels. This highlights the danger of calibrating `trust', as opposed to calibrating the TRBs. Whereas TRB calibration focuses on concrete and measurable behaviors, trust calibration involves influencing something that is directly immeasurable and subject to individual biases when indirect measurements are attempted.

%% file: assurances.tex
An assurance is an AIA property or behavior that can either increase or decrease user trust. The term `assurance' is perhaps earliest used in the context of human-AIA relationships by \citet{Sheridan1984-kx}. \citet{McKnight2001-fa} allude to this kind of feedback in e-commerce relationships as `Web Vendor Interventions'. \citet{Corritore2003-gx} refer to assurances as `trust cues' that can influence how online users trust e-commerce vendors. \citet{Lee2004-pv} discuss `display characteristics', which are methods by which an autonomous systems can communicate information to an operator. More recently, \citet{Lillard2015-yg} provided a formal definition of assurances for autonomous systems that is similar to the one used here. 

\begin{figure}[t]
    \centering
    \includegraphics[width=0.6\linewidth]{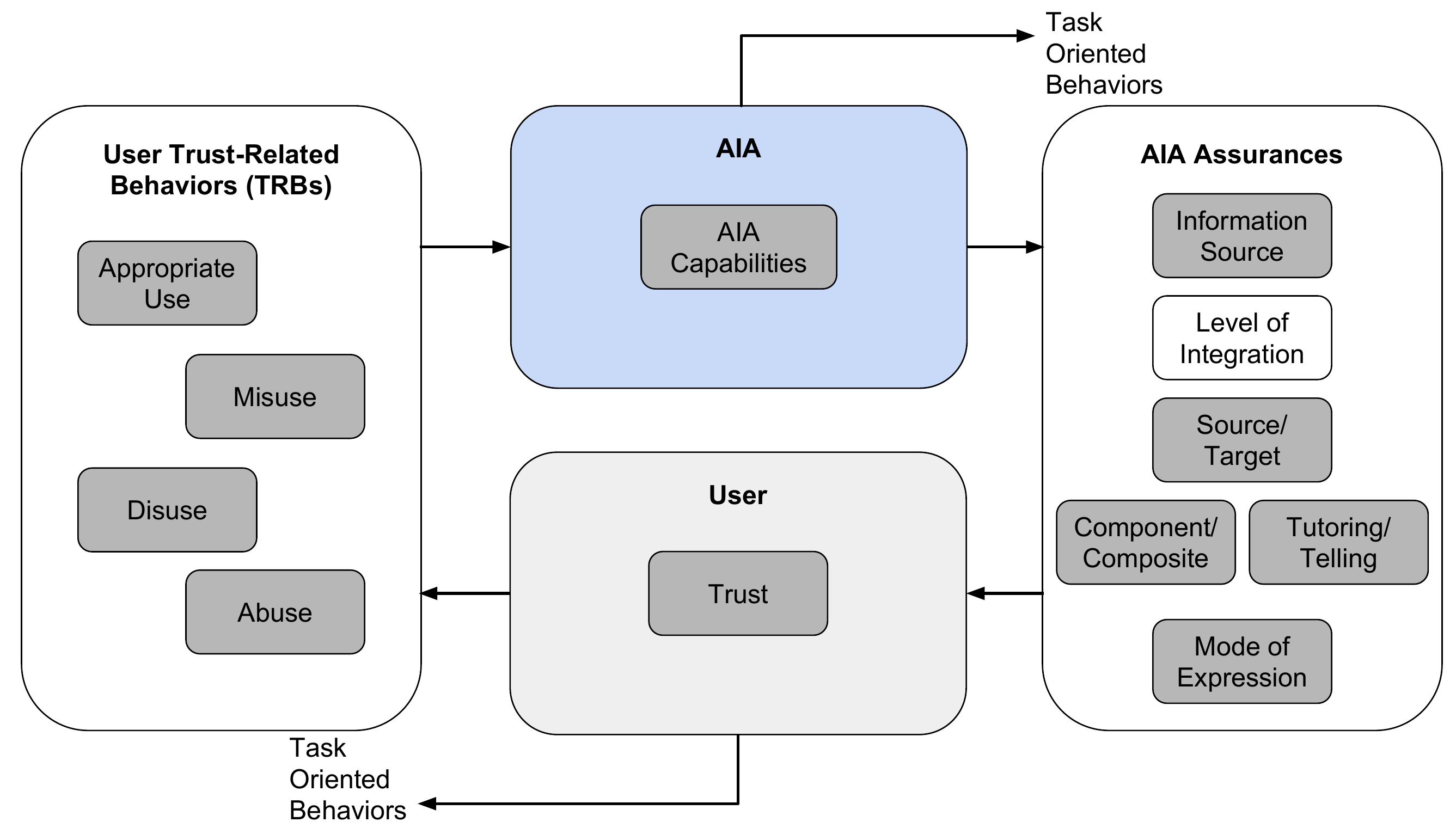}
    \caption{Figure depicting the details of the human-AIA trust cycle.}
    \label{fig:refined_trust}
    \vspace{-0.2 in}
\end{figure}

Assurances can be classified in several different ways. One way to classify an assurance is by its \emph{Information Source:}. Assurances must be informed by some kind of information, whether that means real-time observation of TRBs in order to have feedback, or well accepted concepts of cognitive science as guiding principles of design.
Another approach is to identify the \emph{Source/Target} pair: In a human-AIA trust relationship, assurances link the AIA to the user. The user has multi-dimensional trust in the AIA (see Fig.~\ref{fig:Assurance_classes}), and each AIA capability has multiple dimensions of `trustworthiness'. In designing an assurance it is useful to explicitly identify the source capability, and the target trust dimension (i.e. a certain assurance may have been designed as a `planning-competence' assurance).
An assurance can be considered \emph{Component} or \emph{Composite:} A component assurance stems from one AIA capability to one trust dimension. A composite assurance originates from multiple AIA capabilities to one trust dimension.
Another consideration is whether the assurance is \emph{Tutoring} or \emph{Telling:} An assurance that is dynamic to the different characteristics, and experience of users is a `tutoring' assurance. It is designed to help a user learn, over time, to trust appropriately. Conversely, all other assurances are `telling' in that they are static in regards to separate users.
\emph{Mode of Expression:} Assurances can also be classified by their mode of expression. This includes the method and medium by which the assurance is expressed.
There are many open questions regarding each of these categories; they are discussed further in Sec.~\ref{sec:future_work}, regarding future work.

\emph{Level of Integration:} Herein the `level of integration' of assurances are surveyed. This is useful because it addresses a natural consideration in the design process of AIAs; it also encapsulates well the key approaches that are in use. In this context `integration' refers to the level of effect the assurance has on the core functions of the AIA. As an example: an assurance that, if missing, greatly effects the AIA functionality is considered integral to the AIA. Conversely, a missing assurance that has no effect on the AIA functionality is not integral; we also call this `supplemental'. Between these two extremes there is a natural continuum of integration on which we can classify the different algorithmic approaches to designing assurances; we do so in Sec.~\ref{sec:synthesis}.

%% file: survey.tex
Whereas other researchers have noted the \textit{existence} of assurances, we now directly consider the question: what, exactly, \textit{are} assurances, and how can they be \textit{practically designed} into AIAs? 
This section surveys the related literature to understand what algorithmic approaches can be used to design AIA assurances. 

\begin{figure}[!t]
    \centering
    \includegraphics[width=1.0\textwidth]{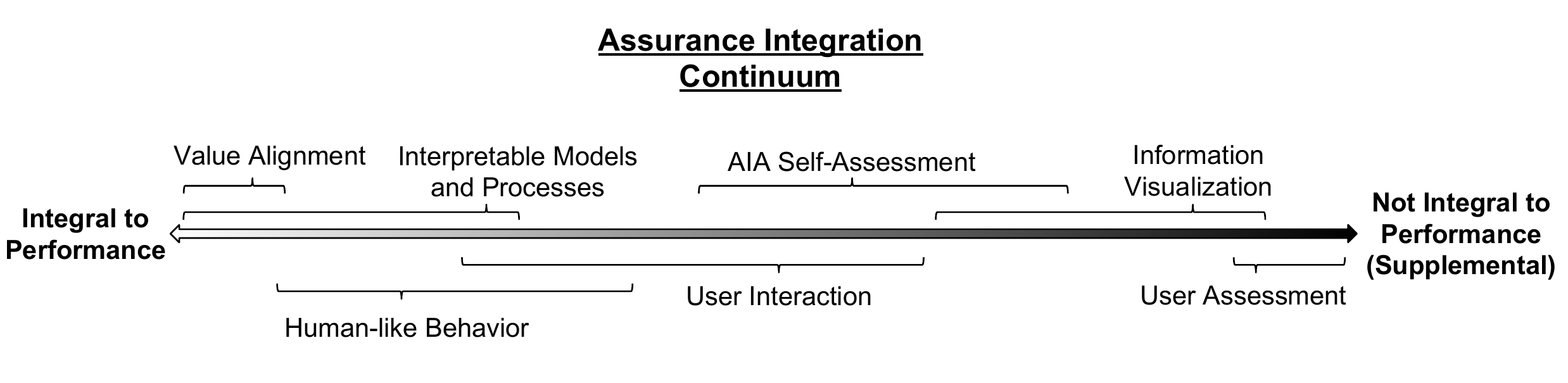}
    \caption{The continuum of the level of integration of algorithmic assurances: to the left are those assurances which are integral to the key functions of the AIA; on the right are assurances which are not integral to performance, i.e. `supplemental' assurances.}
    \label{fig:assurance_continuum}
    \vspace{-0.2 in}
\end{figure}

As discussed in Sec.~\ref{sec:assurances} there are many different ways of classifying assurances. In evaluating different practical approaches to designing assurances we have found that it is easiest to consider the `level of integration' of the assurance in the AIA. The level of integration of an assurance refers to the extent to which the core functionality of the AIA is dependent on the existence of that assurance. Assurances naturally lie on a continuum between being totally integral to the core function of the AIA, and not being integral at all but being generated by artifacts of the underlying task or AIA functionality.  Figure~\ref{fig:assurance_continuum} illustrates this continuum. For simplicity we will sometimes refer generally to assurances as `integral' or `supplemental' based on whether they lie on the left or right side of the figure respectively~\footnote{Note that, while Fig.~\ref{fig:assurance_continuum} shows that the assurance classes occupy large spaces on the continuum, this is not referring to individual `component' assurances. An individual component assurance cannot be both integral and supplemental at the same time; it is located at a point on the continuum. This is not to say that an AIA cannot, simultaneously, have many assurances distributed over the assurance integration continuum, but that these assurances must be considered as separate.}. In the literature we have identified seven main categories for designed assurances that span this continuum.

Practically, understanding the level of integration of different assurances is useful because doing so can indicate at which point different assurances need to enter the design process. For example, an assurance that is integral to the AIA must necessarily be considered from early on in the design process, whereas one that is supplemental can feasibly be added much later. Also, assurances at different levels of integration have similarities in their affects; because of this designers may make different decisions regarding assurance design based on their specifications and goals.

While assurances cannot \emph{guarantee} appropriate TRBs from a user, integral assurances are generally built with the aim of intrinsically guaranteeing---as nearly as possible---certain effects on user trust and TRBs. In contrast, supplemental assurances are typically weaker, and \emph{encourage} appropriate TRBs; they rely much more on the uncertain relationships with human users. Problems can arise, for example, when a designer expects supplemental assurances to have the same effects on TRBs as those of integral assurances. This should generally not be expected. The remainder of this section is dedicated to discussing each of the seven categories in more detail.

\input{methodology.tex}
\input{value_alignment.tex}
\input{interp_models.tex}
\input{humanlike_behavior.tex}
\input{user_interaction.tex}
\input{aia_self_assessment.tex}
\input{visualization_dr.tex}
\input{user_assessment.tex}

%% file: methodology.tex
\subsubsection*{Survey Methodology} \label{sec:methodology}
    While theoretically a two-way trust model could be considered (i.e. in which the AIA also has trust in the user), attention is restricted here to a one-way trust relationship that considers only how user trust (and TRBs) evolves in response to assurances from the AIA. 

    It should be noted that it is practically impossible to perform a fully comprehensive survey of all AIA assurances, due to the broad spectrum of possible assurances, and AIAs in general. As an example, one could rightly argue that control engineers treat metrics like gain and phase margins as assurances for automatic feedback control systems, in much the same way that machine learning practitioners treat training and test accuracy as assurances for learning algorithms---and hence concepts related to robustness, stability, etc. for feedback control systems ought also be included in this survey. Similar arguments exist for assurances developed in fields like econometrics, software testing, aeronautical engineering and many others. While assurances can, in theory, be applied in both the most simple `automatic' systems (like a thermostat), this survey will focus on assurances in more advanced AIAs that make decisions under uncertainty. However, the admittedly narrow scope of this survey does not impede the development of fundamental insights and principles in designing assurances.

    Initially, in order to find applicable research, papers that formally addressed trust, and tried to create models of it, were investigated. This was done with the aim of trying to understand how trust might be influenced. Secondly, literature regarding trust between humans and some form of machine entity was reviewed; this lead to research in fields like e-commerce, automation, and human-robot interaction. Third, research on `interpretable', `comprehensible', `transparent', `explainable', and other similar types of learning and modeling methods were examined. Finally, with that literature as a background, research disciplines investigating computational methods that can be useful as assurances, but in which trust itself is not the main focus, were considered. This information was then used to construct an informed definition and classification of assurances based on methods that are currently in use, or being investigated.
    
    We now proceed to discuss each of the categories from Figure~\ref{fig:assurance_continuum}, starting from the most integral to the AIAs core functionality and proceeding to the least integral.

%% file: value_alignment.tex
\subsection{Value Alignment} \label{sec:value_alignment}
AIAs operate autonomously in delegated tasks, with the expectation that they behave according to users' intent frames. 
Optimization-based algorithms are arguably among the most common and direct approaches for accomplishing this. 
The general idea is to define a \emph{utility function} that normatively governs the AIA's abilities so that desirable behaviors are elicited through maximization of the utility, i.e. such that the AIA behaves rationally in accordance with the user's intent frame. 
A utility function describes the `long-term desirability' of taking certain actions in certain conditions, i.e. beyond immediate benefits or penalties, and should coherently reflect user preferences about the state of the world and AIA behaviors ~\cite{Russell2010-wv}. 
Such mapping of user intent frames to utility functions has two positive benefits. 
Firstly, it ensures that AIA behaviors can themselves be used as assurances: users will tend to trust AIA's more if they are `well-behaved' and acting in accordance with their desired intent than if they are not. 
Secondly, an AIA can generate assurances via auxiliary behaviors that help ensure its utility function is aligned with the user's intent frame.  Since it is practically quite challenging to encode user preferences and intent frames into utility functions, the process of \emph{value alignment}\footnote{Value alignment is more commonly known as `AI Alignment' in AI research~\cite{Yudkowsky2001-hb,Bensinger2014-ul}.} leads to many different algorithmic strategies for generating assurances. 

Consider a generic decision-making problem where an AIA that must make choice $a \in {\mathcal A}$ given some task state $s \in {\mathcal S}$, with scalar utility function $U_A(a,s)$. 
If a user's true utility is represented by scalar function $U_H(a,s)$, then in the ideal situation the AIA seeks the optimal decision $a^* \in {\mathcal A}$ such that, for any $s \in {\mathcal S}$,
\begin{align*}
    a* = \arg\max_{\mathcal A} U_A(a,s) = \arg \max_{\mathcal A} U_H(a,s). 
\end{align*}
Hence, value alignment tries to minimize the difference between the utilities of the AIA and the user. When the utility of the robot $U_A(a,s)$ and the human $U_H(a,s)$ are approximately equivalent (within some tolerance) then the values of the AIA are \emph{aligned} with those of the human. An AIA with aligned values will be considered by users to be more predictable (and thus more competent), because the AIA will be more likely to act in desirable ways. 
\citet{Bostrom2014-fz} provides a well-known example of an AIA whose value is \emph{not} aligned: an autonomous robot is designed, and deployed, with the intent that it make paper clips. To maximize $U_A(s,a)$, the robot then decides to take over the world in order to maximize its resources and ability to make more paper clips. To reasonable human users, this was clearly \emph{not} the intended behavior; the utilities that the robot used for making decisions did not match those that the human must have had. Therefore the robot's resulting behavior was intrinsically an assurance that reduced trust. On the other hand, if the robot were to try to learn from its mistakes and improve (i.e. make $U_A(s,a)$ closer to $U_h(s,a)$) that could be perceived as an assurance that increases trust---the robot can be `forgiven' for making honest mistakes in trying to optimize an ill-posed/under-specified utility function, as long as it is able to recognize and remedy this. 

\subsubsection{Common Approaches:}
There are two algorithmic strategies for value alignment: (i) indirect: approximate $U_H(a,s)$ explicitly via $U_A(a,s)$, and then use this approximation to find $a^*$; (ii) direct: identify $a^* = \arg \max U_H(a,s)$ directly via the use of optimal state-action value functions $Q^*(s,a)$ (which give the utility to be gained if the AIA were to proceed optimally starting from $s$, regardless of its past states or actions). 
These strategies closely resemble techniques used for reinforcement learning problems and their variants (especially inverse reinforcement learning); not surprisingly, most value alignment techniques are rooted in this domain. 
Value alignment research tends to focus on several different issues \cite{Gordon_Worley2018-xy,Amodei2016-xi}; some of the more directly applicable topics and associated methods that point to useful assurance strategies are described below. 
The solutions to these problems are assurances because they afford opportunities for  users to better understand the actual intentions and goals of the AIA, as well as understand how the AIA actually interprets intent frames. 

\paragraph{Reward Hacking and Human-Guided Learning}
The reward hacking problem deals with avoiding and removing unintended consequences in AIA behaviors that arise from imperfections in the specification of $U_A(s,a)$ (as in the paper clip-making robot example above). 
The most popular solution strategies use some form of offline supervisory human guidance or training data feedback in the utility function learning process. 
This approach recognizes the intrinsic difficulty of mapping user preferences to a single scalar utility $U_H(s,a)$ for complex tasks, and leverages sophisticated machine learning and reasoning strategies to identify relevant preferences within $U_A(s,a)$ or $Q^*(s,a)$, depending on the kinds of tasks considered. For instance, one-shot/non-sequential decision-making tasks like image recognition or object perception do not necessarily have dynamical state considerations, but may require potential expansion of the action space for sensible labeling of new object categories. 

In the context of sequential decision making problems, \citet{Hadfield-Menell2017-tl}, \citet{Hadfield-Menell2016-ws}, and \citet{Huang2017-lk} all consider variations of the `inverse reward design' problem using inverse reinforcement learning techniques. In these works, discounted cumulative rewards are used to model utility functions $U_A(s,a)$ and $U_H(s,a)$, where the actual reward factors contributing to $U_H(s,a)$ are unknown but can be inferred from user-generated contextual information at design time. Specifically, \cite{Hadfield-Menell2017-tl} notes that reward factors provided by users in limited training contexts serve as `noisy evidence of intent'. Hence, to avoid situations where an AIA trainee demonstrates desirable behaviors in specific training scenarios but later demonstrates undesirable behaviors in novel scenarios, the AIA must be able to reason over the uncertainty in the user's intent in order to fully capture the context in which it was trained. 
In a different task setting, \citet{Freitas2006-qo} compared two approaches to discovering `interesting' knowledge from large data sets, based on the idea that human users require assistance from complex systems in order to find useful patterns and other interesting insights. He mentions `user-driven' methods that involve a user suggesting interesting templates or providing general impressions in the form of IF-THEN rules. A subsequent comparison to different `data-driven' methods suggests that the latter are not very effective in practice. 
Having said that, user-driven approaches may not fare any better when compared over many users, as each user will likely have different preferences. Other scaled up user-driven approaches, e.g. based on crowd-sourcing~\citet{Chang2017-kl}, can also achieve better accuracy for labeling tasks while also exploring new or ambiguous classes that can be ignored with traditional approaches (especially if training data sets are biased or very limited). \citet{Chang2017-kl} also consider a similar, scaled up, `user-driven' approach called `Revolt' that crowd-sources the labeling of images. It is able to attain high accuracy labeling, while also exploring new or ambiguous classes that might be ignored with traditional approaches. 

Some other methods for designing, learning and eliciting appropriate utility functions are also discussed in \cite{Hadfield-Menell2016-ws,Da_Veiga2012-gh,Garcia2015-rs}.
Despite the differences in AIA application contexts, these methods all provide the user with better context for what \emph{should} be known by system, and for how well it can interpolate/extrapolate. 
These processes allow users to refine their own intent in complex settings, e.g. to reveal or resolve subtle low-level inconsistencies in desired task requirements that would otherwise lead a rational AIA to undesirable behaviors. 

\paragraph{Safe learning and correct-by-construction synthesis:}
In many applications, $U_H(s,a)$ must be safely approximated when certain combinations of $(s,a)$ lead to irreversibly bad consequences. 
Hence, as AIAs try to learn what a user's utility is, they must do so in a safe manner. 
For instance, humans do not learn about the dangers of heights from falling off of skyscrapers. 
Instead we have to do so cautiously over time, and extrapolate from much less drastic experience (i.e. tripping on a curb). 
Safe reinforcement learning (safe RL) methods offer formal strategies and assurances for AIAs to learn in similar ways.
Safe RL has been defined as the process of avoiding ``unintended and harmful behavior that [emerges] from machine learning systems''~\cite{Amodei2016-xi}. Two ways to approach safe RL are: (i) modification of the optimality criterion with a safety factor, and (ii) modification of the exploration process through the incorporation of external knowledge~\cite{Garcia2015-rs}. 

For example, \citet{Lipton2016-dq} design an `intrinsic fear' RL approach that uses a deep Q-network and a `supervised danger model'. The danger model stores the likelihood of entering a catastrophe state within a `short number of steps'. This model can be learned by detecting catastrophes through experience and can be improved over time. \citet{Curran2016-ij}, in a more specific application, asks how a robot can learn when a task is too risky, and then avoid those situations, or ask for help. 
Similarly, \citet{Kahn2017-vy} use Bayesian Deep Neural Nets (using bootstrapping and dropout) to learn about the probability (with uncertainty) of an autonomous vehicle colliding in an environment given its current state, observations, and sequence of controls. Using this model they formulate a `velocity-dependent collision cost' that is used for model-based reinforcement learning. With this approach the vehicle naturally proceeds slowly when there is an elevated risk of collision. This `safety-aware' behavior provides an assurance signal to the user. 

Aside from purely learning-based approaches, we can also consider Validation and Verification (V\&V) methods. 
Not all practitioners are aware that V\&V techniques can generate soft assurances for users. 
This is because V\&V typically refers to the use of formal methods to guarantee the behavior of a system within some set of specifications, which are handed down by a certification authority as requirements to system designers to generate `hard assurances' (formal proofs of the functionality of the system). 
Although these `hard assurances' are not primarily designed for user consumption, they could in principle be exposed to and interpreted for users in certain contexts. 
A prime example is given by \citet{Raman2013-mz}, who developed a formal way for non-expert users to provide structured natural language task specifications to a robot, such that a `correct-by-construction' controller will be built if the specification is valid. 
Otherwise, the robot will provide an explanation about which specification(s) are unrealizable/inconsistent and will cause failure. 
In the context of a practical self-driving car application, \citet{Ghosh2016-dl} presents a framework called Trusted Machine Learning (TML) for learning models from dynamically generated data that fit pre-determined `trustworthiness' constraints. 
These approaches are promising in that they not only present a way to communicate when and why specified tasks cannot be performed or certain actions cannot be taken, but also provide positive assurances in the form of guaranteed, formally verified, AIA processes for performing desired tasks (plans, models, etc.). 
While this directly addresses the competence and predictability components of AIA trust, the `raw' expression of these assurances does not formally account for effects on user trust or TRBs in formulating explanations. 

\paragraph{Robustness to context shifts}
How can an AIA determine when the basis and provenance of its approximation to $U_H(s,a)$ or $Q^*(s,a)$ is no longer valid for a particular task? 
This problem has attracted much recent attention in the learning literature under the guise of `nonstationary' learning. 
Nonstationarity refers to the complex challenge of training a model based on data from one distribution $D$ while taking into account that the test distribution $D^\prime$ will likely shift through time~\cite{Quinonero-Candela2009-fj}. 
For instance, in the context of classification problems, \citet{Sugiyama2013-ci} propose using importance sampling Monte Carlo to formally detect events related to `covariate shift' (training and test input data follow different distributions) and `class-balance change' (where the class-prior probabilities are different in training and test phases, but where there is no covariate shift). 
Similarly, \citet{Charikar2017-kr} address learning from `untrusted' data, which could be subject to adversarial attack or unknown nonstationarity. 

These methods can be more generally adapted and developed beyond learning tasks, in order to evaluate the sensitivity of as-designed AIA's capabilities to possible changes in task context not captured/considered at design time (an example of coping with `unknown unknowns').  
If the sensitivities imply a significant deviation in $U_A(s,a)$ or $a^*$ from expected values (i.e. from user intent frame as initially understood), or indicate the presence of new $(s,a)$ pairs that are not accounted for by $U(s,a)$ (e.g. test data that is very far from the training set), then the AIA can inform the user accordingly and thus possibly opt out of performing tasks that are now potentially `out of scope'. This provides direct low-level behavioral assurances about changes in predictability, competence, and situation normality, though these may not be immediately understood by non-expert users. 

\subsubsection{Grounding Example:}
In the case of the `VIP Escort' problem (described in Section~\ref{sec:mot_example}), value alignment might be used as an assurance in the following way, starting with the assumptions that:

\begin{itemize}
    \item The UGV has just begun an attempt to escape the road-network
    \item The UGV uses safe RL to learn its escape policy
    \item The operator is able to observe the UGV during its entire escape attempt
\end{itemize}

The operator has used several different UGVs for similar tasks. This newer model uses `safe RL' to learn its policy. When observing the UGV's attempt at escape the operator notices a difference in how the UGV operates. Whereas the older UGV models would sometimes do risky things, this UGV seems to navigate dangerous situations much better. 

\paragraph{\textbf{Discussion of Example:}} In this case, safe RL enabled the UGV to treat situations that an operator might classify as `dangerous' with more care. With this integral capability the UGV assures the operator that it is more competent.

%% file: interp_models.tex
\subsection{Interpretable Models and Processes} \label{sec:interp_models}
Another way to provide assurances about AIA conformance to user intent frames is to expose the models and algorithmic processes governing its actions directly to the user. If these models and processes also happen to be easy for users to interpret, then the user can (ideally) acquire a well-formed and highly predictive `theory of mind' for the AIA's behavior, with little or no effort . 
\citet{Doshi-Velez2017-xy} give an argument for why interpretability is critical in AIA systems since interpretability `is used to confirm other important desiderata of [machine learning] systems'. 
Yet, perhaps unsurprisingly, `interpretability' and the attendant desiderata still elude formal universally accepted definitions. 
They also use the words `interpretable' and `explainable' interchangeably. In contrast, we treat them as distinct descriptors. We discuss models that are inherently interpretable here, and models that can be understood by explanation in Section~\ref{sec:reduce_complexity}.  The difference is that interpretability (in our view) implies that the actual process/model used by an AIA is self-explanatory, whereas explainable models can be made interpretable by post hoc operations but do not necessarily explain the actual model/process used by an AIA. 
Being able to interpret the actual model/process used by an AIA helps human users to more appropriately understand their behaviors, and thus exhibit appropriate TRBs in turn. This approach to assurance also captures broader AIA processes and models that rely on rules, heuristics, etc., rather than just those that rely on optimization of some particular utility. 

\subsubsection{Common Approaches:}
Two main approaches to designing interpretable AIA models and processes are considered here. 
The first is to \emph{assess an existing set of candidate models/processes} in order to evaluate their interpretability in the context of a particular task, and then select the best candidate. 
This is typically done with certain classes of models or solution processes, e.g. whether to use decision trees vs. decision tables for a given planning task. 
The second is to \emph{synthesize interpretable models/processes} by leveraging human designer input during the model/solution-building process. 
The first approach requires pre-defined measures of interpretability, and thus some mechanism for capturing ability to gain insights into competence, predictability, and situation normality. This also presupposes that the candidates are inherently interpretable along these lines to begin with, which may rule out methods that perform well on certain tasks. 
The second approach allows designers to apply domain knowledge to determine metrics for interpretability, although this can lead to solutions that do not perform as well as those that are less interpretable. 

What is the assurance mechanism that potentially leads to proper TRBs in either approach? 
Essentially, allowing the user to access and examine an interpretable model/process also allows them to simultaneously assess competency, predictability, and situational normality components of trust. If the models/processes are perfectly interpretable, then a user could understand exactly how the AIA would perform its task (i.e. down to a mechanical/programmatic level). 
This gives the user a `mental model' of what the AIA would consider to be situational normality and how AIA would respond in different situations (predictability and competence). 
The caveat here is that incorrect TRBs may arise if the user mis-interprets or only understands part of the model/process. 
This is a significant risk in highly complex or specialized problems, where users may not actually have sufficient training or expertise. This also poses concerns for how/when users can access interpretable AIA components. 
Unlike value alignment (where user accesses assurances only through behavior of AIA itself), the user has more freedom in deciding when and how to `peek under the hood'. This relates to assurances based on information visualization discussed later, except that here the information being given to the user are the actual AIA algorithms themselves, as opposed to byproducts or after effects of those algorithms.

\paragraph{Assessing Interpretability:}
\citet{Van_Belle2013-ph} suggested three ways to ascertain the level of interpretability and potential utility of learned models (compare to categories proposed by \citet{Lipton2016-ug}): 1) Map them to domain knowledge; 2) Ensure safe operation across the full operational range of model inputs; and 3) Assess whether important non-linear effects are accurately accounted for. This work identifies certain strengths and weaknesses of different techniques, but ultimately concludes that no method is clearly best in all situations. 
Along similar lines, \citet{Huysmans2011-th} compared decision trees, decision tables, propositional if-then rules, and oblique rule sets to understand which set of methods is `most interpretable'. It was experimentally determined that decision trees and tables tend to be easier to interpret, but it is noted that each method could perform better than others in different applications. For example decision trees and tables are typically better suited for answering a symbolic question (which requires a local understanding of a model) like: \emph{how does the model classify observation $X$'?}. This is in contrast to a spatial question (which requires a global understanding of the model) like: \emph{is it correct that applicants with a high income are more likely to be accepted than applicants with a low income?}
Having quantified the interpretability of a model given different classes of problems, and different requirements of users the appropriate model can then be selected during design to fit the needs of a specific application.

\begin{figure}[htbp]
    \centering
    \includegraphics[width=0.45\textwidth]{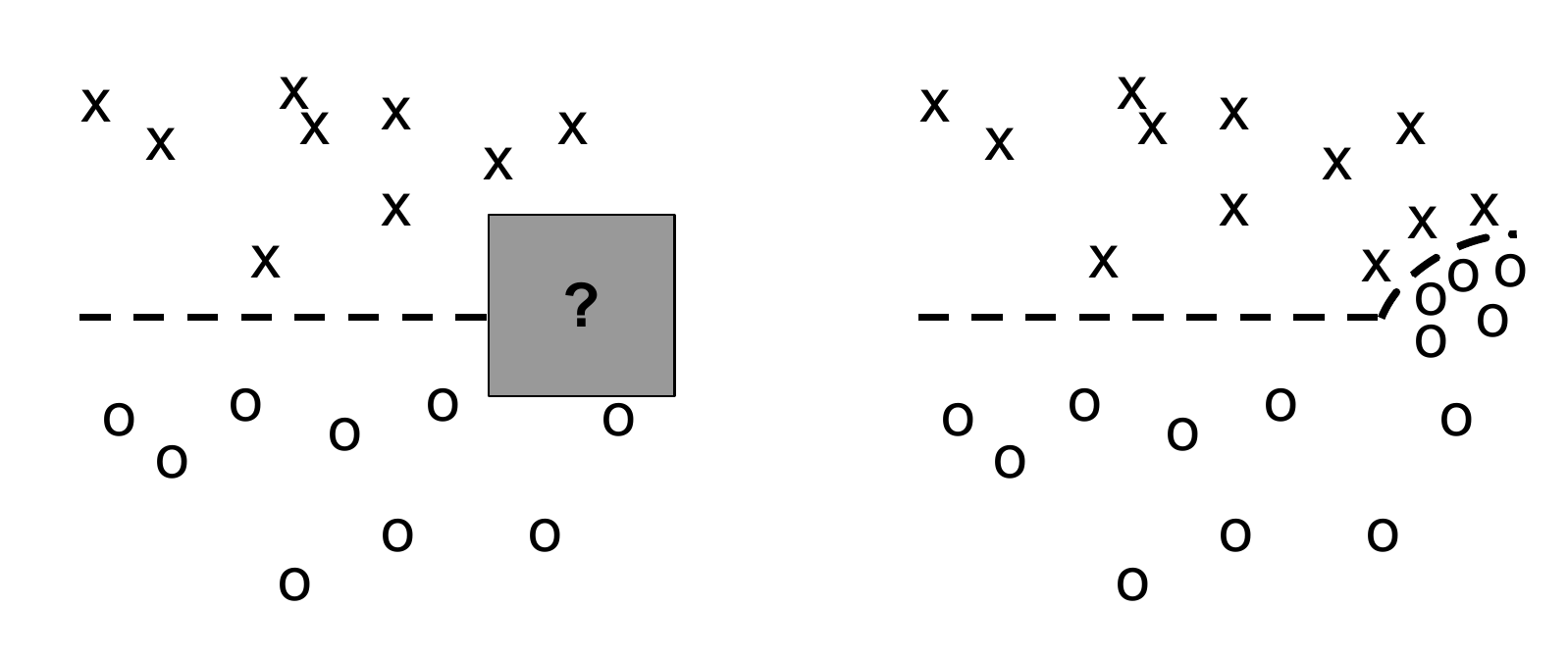}
    \caption{Example of simple global interpretable learning model on the left, and on the right a more complex locally interpretable learning model that can be used when more precise understanding of a specific decision made by the learner is required. }
    \label{fig:ruping}
    \vspace{-0.1 in}
\end{figure}

\paragraph{Interpretable Model Synthesis:}
\citet{Ruping2006-xj} asks how classification results and the accuracy-interpretability trade-off can be made more transparent to those who design and use classifiers. He explores one approach by combining simpler global models with more complex local models that are built around learning results (\citet{Otte2013-oo} and \citet{Ribeiro2016-uc} implement similar ideas as well). 
Figure \ref{fig:ruping} illustrates this idea. 
The explanation of Fig.~\ref{fig:ruping} could be something like: `The classification boundary is generally a horizontal line. However, for a small region on the right hand side the boundary is shaped roughly as an inverse quadratic starting from the horizontal line'.

Considerable effort has also gone into endowing `grey box' and `black box' models with interpretable features. 
For instance, \citet{Abdollahi2016-vn} investigate making collaborative filtering models more interpretable by using a conditional restricted Boltzmann machine (RBM). \citet{Ridgeway1998-lv} use `weight of evidence' (WoE) as a boosting method that is more amenable to interpretation, and show that the performance WoE is on par with AdaBoost. \citet{Choi2016-by} construct a recursive attention neural network to remove recurrence on the hidden state vector, and instead add recurrence on the visits of patients to doctors, as well as on different diagnoses during those visits. In this way the model is able to predict possible diagnoses in time, and a visualization can be that that indicates the critical visits and diagnoses that lead to that prediction.

Learning of human-understandable representations for data and feature selection also provides another avenue for developing assurances  \cite{Bengio2013-uv, Guyon2003-fj}. For instance, \citet{Mikolov2013-lt} studied how to represent words and phrases in a vector space for natural language text learning; this enables simple vector operations for understanding word sense similarity and relative relationships learned from text corpora. For example, the vector addition operation \emph{airlines+German} yields similar entries that include \emph{Lufthansa}. Such representations encode knowledge that can be easily checked and understood by humans, and thus implicitly facilitate interaction and calibration of trust (see \cite{Haury2011-zi} for another example). The problem of discovering human understandable features and representations in more general settings still remains an open question. Currently, the main question for representation learning is how to find the `best representations' for a particular application---not necessarily the representations and features that are `most humanly understandable'. This is not surprising, since human-understandable representations and features are not necessarily optimal for the criteria that AIAs are typically designed against. 

Contrary to the belief that interpretable models are necessarily worse performing than their less interpretable counterparts, several researchers have shown that this is not always the case (at least in the context of machine learning). However, the real trade-off is the amount of work that goes in to crafting the interpretable model from the start; these methods are often custom designed for certain tasks and are not easily transferable to other problems. Because of this, AIA designers must strike a balance between \emph{interpretable models}, \emph{explainable models}, and \emph{black-box models}.

\citet{Park2016-ld} point out that real interpretability in complex tasks still requires expert knowledge to make sense of complicated features; in essence: \emph{people are needed at both ends of interpretable models}. For instance, \citet{Jovanovic2016-gw} use `Tree-Lasso' (TL) logistic regression with domain knowledge (i.e. medical diagnostic codes) to group similar conditions, and then use TL regression again on that information to develop a sparser model. \citet{Zycinski2012-jj} also use domain knowledge to structure a data matrix before feature selection and classification. See also \citet{Zhang2018-no,Khoa2018-gh} for other related examples. 
This kind of approach is also illustrated by those who use those who use `theory guided data science' (TGDS~\cite{Kumar2016-yw,Faghmous2014-og}). As one example \citet{Morrison2018-fz} address the situation where an imperfect analytical model is available for chemical reaction kinetics: the theoretical reaction equations are well known, but a `stochastic operator' is added on top of this to account for uncertainties and modeling errors. In adopting this approach the model becomes interpretable (to experts).

\subsubsection{Grounding Example:}
In the case of the `VIP Escort' problem (described in Section~\ref{sec:mot_example}), interpretable models might be used as an assurance in the following way, starting with the assumptions that:

\begin{itemize}
    \item The UGV has just begun an attempt to escape the road-network
    \item The UGV is using a decision-tree for selecting different movements
    \item The operator is able to view the decision-tree model the UGV is using
\end{itemize}

While the operator is monitoring the progress of the UGV in its attempt to escape the road-network they are able to consult the decision-tree model. In this case the operator chose to consult the table when they saw the UGV make an unexpected turn at a given intersection. The operator identified the conditions that led to the decision and found that the UGV was not well equipped to execute the decision the operator thought was best.
\paragraph{\textbf{Discussion of Example:}} In this example the use of a decision-tree as a model enabled the operator to investigate unexpected behavior. During inspection they identified certain conditions that led to a decision, and they found that the UGV was not \emph{competent} to perform what the operator thought was a better decision. Because of this the operator better understood the decision the UGV made, and will have a more appropriate level of trust in future interaction.

%% file: humanlike_behavior.tex
\subsection{Human-Like Behavior} \label{sec:human_behavior}
Since humans are accustomed to forming and evaluating trusting relationships with each other, imitation of human-human communication and interactive behaviors provides yet another avenue for developing AIA assurance strategies. 
Support for this idea is given by \citet{Tripp2011-rx}, who compared human trust in other humans against human trust in intelligent interactive technology. 
They found that, as the technology becomes more `human-like', self-reported levels of trust in technology become more similar to levels of trust in other humans.

\citet{De_Visser2018-kd} also specifically discusses different methods by which AIAs can be more human-like in order to `repair trust' with users (here, trust repair is roughly analogous to assurances, but focuses on re-building trust after it is lost). Among several other possibilities, they suggest that an AIA might repair trust by anthropomorphizing (responding using a human communication channel), or by explaining its actions in the same way a person would. 
Such human-like behavior opens the door for AIAs to exhibit `non-rationally motivated' behaviors (i.e. suboptimal, as opposed to irrational actions), if these conform to social norms or other psychological cues that provide useful assurances about predictability (e.g. a robot arm that executes legible motions), competency (e.g. a robot which slowly backs away from unfamiliar or potentially dangerous objects), or situation normality (e.g. a robot car that apparently rubbernecks near an unfamiliar scene on the road). 

\subsubsection{Common Approaches}
Generally, we do not have algorithms that describe how humans interact with each other (yet), and must settle for heuristics or best attempts to create human-like behavior via algorithms. From a high level, researchers have addressed these: nonverbal communication, and mannerisms.

\paragraph{Nonverbal Communication:} 
Nonverbal communication can take many different forms. One popular approach is to use motion or gestures. \citet{Szafir2014-ok} investigated how to enable `Assisted Free Flyer' robots (quad-copters that are made to interact with humans in close spaces) to communicate by using gestures. In doing so they use `motion primitives' (a basic vocabulary of movements) that were inspired by basic `character animation' principles \cite{Van_Breemen2004-rz}. 
In their evaluations of these primitives with human participants in the presence of free flyers, they found that human users significantly found the free flyers to be more natural, and felt safer around them. Later \citet{Szafir2015-iy} also experimentally showed the effectiveness of using illuminated `turn signals' and pairs of human-like `eyes' that shifted with free flyer heading (much as human eyes do when people walk in a crowd) to help users more easily interpret the vehicle's intended movements and actions. These works provide strong support for `commonsense communication' assurances aimed at predictability in physical user-AIA interactions (even if indicators like `moving eyes' do not actually see anything). 
Likewise, \citet{Dragan2013-wd} investigate `legible motion planning', i.e. planned robotic physical movements and gestures that, by themselves, convey intended actions and goals. For example, a table-setting robot may grasp a plate on both sides from the top using two end effectors if it intends to shift the position of the plate along the table surface, whereas it may grasp the same plate with only one hand from the side if it intends to pull away and remove the plate from the table. 
Legible motion is used by humans working in close proximity, and so can also be useful and important for situations in which a physically embodied AIA and person are collaboratively working in close proximity to each other. Similarly, in more recent work \citet{Kwon2018-xt} investigates calculating trajectories that convey `incapability', which is \emph{what} the AIA is trying to do, and \emph{why} it is unable to do so. 
See also \cite{Admoni2016-db} for related work. 

\paragraph{Mannerisms:}
Humans are naturally inclined to leverage social interaction cues and adherence to/violation of social norms as evidence for assessing the trustworthiness of other humans in everyday interactions. 
AIAs can leverage these inclinations to provide simultaneous assurances of their competence, predictability, and situation normality. 
Consider, for instance, a recent `mini-Turing Test' example from the popular media: at Google/IO 2018, Google Duplex \cite{Google2018-eb} was introduced through a demo where it placed a phone call to make a reservation. 
An oft-remarked feature of this demo is the great difficultly (if not near impossibility) of detecting whether or not the Duplex voice is human -- down to the words spoken, tone of voice, and speech mannerisms (which included `um\ldots', pauses, and shortened sentences). 
The human on the other end of the call was none the wiser, and trusted that they were in fact speaking to a regular human customer -- when in fact they were speaking in a completely natural manner to the product of a recurrent neural network (RNN) trained on anonymized phone conversation data. 

More formally, \citet{Salem2015-md} investigated the effects of autonomous task errors, task types, and `system personality' on cooperation and trust for humans who observed a domestic robot performing house tasks, such that the robot implicitly showed competence by its mannerisms and successes/failures during tasks. In this case, the mannerisms and competency of the robot were completely under control and hard-coded into the system. Regardless, when participants were asked to cooperate with the robot on certain other tasks, the strange/unexpected operation of the robot was found to influence the self-reported trust levels of the participants.

\citet{Wu2016-ei} investigated how a person's decisions in a coin entrustment game are affected by their belief in whether they are competing against an AIA or another human player (which, unbeknownst to participants, was in fact an AI with some programmed human-like idiosyncrasies, e.g. variable wait times between turns). Trust in this context was measured directly by the number of coins a participant was willing to lose by putting them at risk to the other player. The experiment found that the participants trusted the AI opponent more than they trusted the `human' opponent; the authors suggest that this may be due to the perception that the AI opponent did not have feelings and operated in a more predictable and consistent `machine-like' way. Given that the `human' was an AI as well, this experiment illustrates that `machine-like' behavioral consistency can lead to implicit positive effects the trust of the participant in certain contexts.

\subsubsection{Grounding Example:}
In the case of the `VIP Escort' problem (described in Section~\ref{sec:mot_example}), human-like behavior might be used as an assurance in the following way, starting with the assumptions that:

\begin{itemize}
    \item The UGV is about to begin an attempt at escaping the road-network
    \item The operator can observe all the actions of the UGV via video feeds at intersections
    \item The UGV has been designed with the ability to use gestures in order to indicate its `incapability' as in \cite{Kwon2018-xt}
\end{itemize}

As the UGV begins the escort problem, the human supervisor is monitoring progress. When the UGV reaches a certain intersection of the road network the supervisor expects the UGV to take a path $A$, but it does not. However, before choosing to take path $B$, the UGV made a movement that, to the operator, indicated that it considered attempting to traverse $A$. Due to the attempt the supervisor was able to surmise that the UGV wanted to take that path but couldn't due to some limitation.

\paragraph{\textbf{Discussion of Example:}} In this case the UGV is able to maintain appropriate trust of the supervisor because the supervisor was able to interpret the `gesture' that UGV was using. This highlights the assuring effects that human-like communication/behaviors can have on users.

%% file: user_interaction.tex
\subsection{User Interaction} \label{sec:user_interaction}
Despite the oft-repeated sentiment that advanced AIAs will `soon' be able to operate with little or no human involvement, those who have more practical experience with AIAs are much more skeptical of this claim, and point out that it is highly unrealistic to expect AIAs to ever function `perfectly out of the box' with true total autonomy \cite{Bradshaw2013-ck}. 
A popular and promising avenue for surmounting the inevitable shortcomings of AIAs, and thus engendering trust in users, has therefore been to put the users `in-the-loop' (or `on-the-loop') as collaborative partners who can augment (or supervise) AIA capabilities. 
In formulating algorithms for AIA capabilities that leverage user inputs, the user becomes analogous to a supervisor working alongside those they supervise; in doing so they are able to provide useful feedback in real-time, lend their expertise, and better appreciate the decisions and outcomes of the team's work. Such collaborative problem solving not only gives users a chance to directly assess AIA competence and predictability through experience (assurances), but also provides a way for users to continuously engage AIAs in accordance with their actual capabilities (appropriate TRBs). 
Note that user interaction techniques are \emph{not} the same as user assessment techniques discussed later, since user assessment techniques do not involve fundamentally changing AIA algorithms or capabilities to exploit user interaction.

\subsubsection{Common Approaches:} 
Users can be exploited to provide or augment any of the AIA capabilities in Fig. \ref{fig:AIcapabilities} on many different levels. 
At one extreme, a user might fully replace or augment a subset of core AIA capabilities, e.g. to act as a high-level `sensor' and planner for an autonomous robot in a navigation task \cite{Kaupp2008-yr}.  
On the other extreme, the human might have a very weak involvement in the core perception functionality of an AIA, e.g. to validate the labeling of image data. 
Since the literature in this area is quite vast and ongoing research quite active, we focus here, for the sake of brevity, only on a few typical methods from the human-robot interaction literature where the AIA (an autonomous robot) engages the user as an additional `sensor'/perception agent or `controller'/planning agent. The references cited in these works also point to a host of other related and relevant techniques, which in turn can (and have) been adapted to other AIA capabilities such as learning, reasoning, knowledge representation, etc.

\citet{Sweet2016-dw} investigate the use of  humans as `soft' sensors for target localization tasks, whereby semantic natural language observations (`Target is by the bridge', `Nothing in the street') can be directly combined with conventional `hard' robot sensor data (from cameras, lidar, sonar, etc.) in order to improve and augment the robot's Bayesian state estimation algorithms. 
They apply their approach in a scenario called `Cops and Robots' where a single `cop' robot tries to locate mobile `robber' robots in a semantically rich indoor environment. 
In this case the human acts as a `deputy' that remotely interacts with the system. The human can see security camera footage of the building in which the cop is searching, and can offer natural language feedback to the cop robot when appropriate. If the human offers information, it can be fused into the cop robot's estimation model, but in the meantime the cop robot operates autonomously to plan its motion without human assistance. 
Along similar lines, \citet{Kaupp2008-yr} empirically identify the appropriate level of autonomy for a robotic navigation system while taking into account the amount of sensory interaction required by a human supervisor. In this case the robot has sensors of its own, but can also ask for user input when the value of information (VOI) is high enough (i.e. is it worth asking a human sensor for information given that there is a cost?); they define the threshold VOI by performing human trials before deployment of the system in order to optimize the involvement of the human user.

\citet{Tellex2014-uc} consider planning algorithms that are augmented by human natural language commands for an autonomous assembly robot that can detect when it has failures (conditions that don't match expectations based on internal models). When this occurs the robot requests help from the human user to resolve the problem. In this way the human and robot are dependent on each other to accomplish a task. Since the user knows that, if needed, the robot will ask for help, they can more appropriately trust that unknown problems won't occur without them being informed.
\citet{Freedy2007-sg} studied performance measures for  mixed-initiative human-autonomous robot teams (where users and robots share planning and decision authority), and examined the extent to which such teams can only be successful if ``humans know how to appropriately trust and hence appropriately rely on the automation''. They explore this idea by using a tactical reconnaissance scenario where human participants supervised an unmanned ground vehicle (UGV)  platoon with three levels of autonomous targeting/firing capability (low, medium, high); these levels were dependent on the experimental conditions. The operator needed to monitor the UGV in case it couldn't perform as desired; in such cases the operator could intervene to resolve the problem. Operators were trained to recognize signs of task failure, and to only intervene if they thought the mission completion time would suffer. 

\subsubsection{Grounding Example:}
In the case of the `VIP Escort' problem (described in Section~\ref{sec:mot_example}), user interaction might be used as an assurance in the following way, starting with the assumptions that:

\begin{itemize}
    \item The UGV has just begun an attempt to escape the road-network
    \item An interface system exists by which the operator can receive and provide information to the UGV
\end{itemize}

The UGV is capable of operating autonomously, but also can benefit by asking for assistance or information when necessary, e.g. using a natural language interface for augmented planning and sensor fusion. In this way the functionality of the UGV can be greatly improved via interaction with the user. As the user interfaces with the UGV and is able to provide feedback and information about the best known location of the pursuer based on information unavailable to the UGV they have more trust in the competence, predictability, and situational normality of the UGV.

\paragraph{\textbf{Discussion of Example:}} In this scenario the user is more immersed in the functioning of the UGV. Not only are they able to respond to queries from the UGV, but they can also provide direct observations as well. Subsequently, the user feels more immersed in the functioning of the UGV and is more cognizant of appropriate TRBs.

%% file: aia_self_assessment.tex
\subsection{AIA Self-Assessment} \label{sec:aia_self_assessment}

The techniques of previous sections generally tend to provide integral assurances (i.e. designed as part of core functionality of AIA capabilities) that are artifacts of interactive algorithms designed to compensate for shortcomings in AIA capabilities. This section focuses on `introspective' assurances that inform users of competency limits and boundaries of AIA capabilities without requiring user interaction, and that can generally be separated from core AIA functionality (i.e. without requiring modification of core, underlying, AIA design).  These self-assessments can provide users with insights regarding either or both of the following related issues: (i) what information and tasks are actually within the AIA's reach?, and (ii) what is required by the AIA to actually do its assigned task? 
In contrast to user interaction techniques: the analogy here is of a subordinate telling a supervisor what she is/is not capable of, or telling the supervisor what she would need in order to carry out specific task at hand to achieve a specific outcome, or what the possible outcomes actually would be for that specific task. 

\subsubsection{Common Approaches:}
The literature in this section can be split into two high-level categories. 
The first set deals with how an AIA can algorithmically account for its uncertainties in its models of its task, environment, operating context, and capabilities. 
These kinds of assurances help inform the predictability and situation normality aspects of trust. 
The second set of methods attempt to algorithmically reduce complex `uninterpretable' models or processes that underlie AIA capabilities into more interpretable ones by providing explanations. 
Here the AIA makes an active attempt at processing data and making information available to the user to inform the competency aspect of trust. 

\input{quantify_uncertainty.tex}
\input{reduce_complexity.tex}

\subsubsection{Grounding Example:}
In the case of the `VIP Escort' problem (described in Section~\ref{sec:mot_example}), self-assessment might be used as an assurance in the following way, starting with the assumptions that:

\begin{itemize}
    \item The UGV is about to being an attempt to escape the road-network
    \item The UGV is using the `solver quality' metric mentioned by \citet{Aitken2016-fb}
    \item The operator has access to an interface where they can view the self-confidence metric calculated by the UGV
\end{itemize}

Before the UGV begins its attempt it is able to assess its `solver quality' given the specific, previously unseen, road-network based on similarities between the current network and ones that it has encountered before (i.e. problem features that are important to determining the quality of the approximate solution produced by the policy). The UGV reports that it has high confidence in its solver quality, and the operator is assured that they can trust the solver in this situation.

\paragraph{\textbf{Discussion of Example:}} In this case the UGV is able to assure the operator of the quality of the solver in the specific road-network. Generally, the UGV reduced what could be a very complex analysis into a simple format for the operator to interpret. This is in contrast to the operator viewing policies, models, algorithms, and complex probability distributions.

%% file: quantify_uncertainty.tex
\paragraph{Accounting for Uncertainty:} \label{sec:acct_uncertainty}
An AIA that can predict its performance on different tasks can provide assurances about competence, predictability, and the situational normality of a given task. Several researchers have worked to improve this ability in visual classification \cite{Zhang2014-he,Gurau2016-hs,Churchill2015-ei,Kaipa2015-hy}. 
For example, to ensure that visual classifiers don't fail silently in novel scenarios, \citet{Zhang2014-he} learned models of errors on training images to predict errors on test images. \citet{Kaipa2015-hy} consider 3D visual classification of assembly line parts for robotic pick and place tasks, and develop statistical goodness-of-fit tests to estimate the likelihood that robots can use their sensors to find parts matching desired ones. These approaches allow the AIA to assess capability and present appropriate assurances to users, though without any formal notions of trust. 

\citet{Mitchell2018-jw}, discuss, in the context of a `never ending learning problem' (i.e. where the AIA perpetually learns over time), how an agent can quantify uncertainty on unlabeled data given three requirements: 1) three or more approximations of a function are available, 2) the assumption that these functions are more accurate than chance, and 3) these functions have independent errors. The rates at which these functions agree on classification of unlabeled examples can be used to solve for their exact accuracies. Doing this allows the system to actively reduce uncertainty, by seeking relevant data.
In the context of image classification, \citet{Paul2011-vr} introduced `perplexity' as a metric that represents uncertainty in predicting a single class and is used to select the `most perplexing' images for further learning. There have also been several attempts to use Gaussian processes (GPs) to actively learn and assign probabilistic classifications \cite{MacKay1992-sp,Triebel2016-kj,Triebel2013-ow,Triebel2013-ku,Grimmett2013-gj,Grimmett2016-yc,Berczi2015-rd,Dequaire2016-kh}. As with perplexity-based classifiers, the key insight is that if a classifier possesses a measure of uncertainty, then that uncertainty can be used for efficient instance searching, comparison, and learning, as well as reporting a measure of confidence to users. The key property of GPs to this end is their ability to produce output confidence/uncertainty estimates that grow more uncertain away from the training data. This information can be readily assessed and conveyed to users, even in high-dimensional problems. This property has also found much use in other AIA active learning problems, e.g. Bayesian optimization \cite{Snoek2012-tt, Brochu2010-tj,Israelsen2017-zb}.

Neural network (NN) models are commonly considered black-box models, and methods to represent uncertainty have not historically been available. However, there have been several recent advances to make this possible to some extent~\cite{Gal2016-om,Gal2016-eq}. Bayesian neural networks (BNNs) are a method by which we can draw insight about the uncertainty of a neural network's predictions; this is possible by placing prior distributions over the weights in a NN. \citet{Kendall2017-ry}, in the context of computer vision, also use deep BNNs to help visualize epistemic (input) and aleatoric (model) uncertainty for each pixel of an image. 
Similarly, \citet{Kahn2017-vy} use deep BNNs to learn about the probability (with uncertainty) of an autonomous vehicle colliding in an environment given its current state, observations, and sequence of controls. Using this model they formulate a `velocity-dependent collision cost' that is used for model-based reinforcement learning. 
In order to help predict uncertainty in real-time robotic applications that learn from demonstrations, \citet{Choi2017-th} use mixture density networks (MDNs)---neural networks that learn parameters of a Gaussian mixture distribution---to model complex distributions from human demonstrations.

Models and logic are not trustworthy by themselves; they may be flawed to begin with, or become invalid when assumptions or specifications are violated. Thus, there is great interest in providing assurances that the models and assumptions underlying different AIA processes are in fact sound. \citet{Laskey1991-mf}---with the intention of communicating model validity to users of `probability-based decision aids'---notes that it is infeasible to perform a decision-theoretic calculation to determine if model revision is necessary. She presents a class of theoretically justified model revision indicators, based on the idea of constructing a computationally simple alternate model and then initiating model revision if the likelihood ratio of the alternate model becomes too large (see also \citet{Zagorecki2015-qy,Habbema1976-xd}). \citet{Ghosh2016-dl}  present `model repair' and `data repair' strategies that can be used when the current model does not match the observed data, at which point the model and data can be repaired, and control actions can be replanned in order to conform with the formal method specifications. One challenge is how the `trustable' constraints should be identified, as this places a strong burden on the certifying authorities and system designer to foresee all possible failures.

%% file: reduce_complexity.tex
\paragraph{Reduce Complexity} \label{sec:reduce_complexity}
Representations within an AIA are often complex. Sometimes using inherently less complex, `interpretable models' (as discussed in \ref{sec:interp_models}), is the most straight forward way to address this challenge. However, in some cases it is desirable to maintain complex, less interpretable representations (e.g. for performance reasons) and then reduce the inherent complexities (possibly post-hoc) to aid human users.

One typical approach is to generate explanations, but how should explanations be provided? There are also considerations regarding whether explanations should occur by two-way interaction between system and user, by natural language interaction, or by probabilities. Some of the answers to these questions lie more in the realm of cognitive science. Still, natural language and other communication modalities could be used~\cite{Hayes2017-nt}. Specifically, \citet{Olah2018-rp} investigate how predictions of NNs can be explained through visualizing how different parts of the network respond to certain images. They propose combining several different approaches to get a holistic view of the NN behavior. Specifically, they use feature visualization (what a neuron is looking for), and attribution (how it affects the output).

There are several classes of explanations. \citet{Abdollahi2018-uw} propose three in the context of collaborative filtering: `neighbor style' (explanation based on examples from similar situations), `influence style' (present the most influential items that led to a certain model output), and `keyword style' (identify common features between user keywords, and content). \citet{Otte2013-oo} and \citet{Ribeiro2016-uc} implement analagous ideas in the realm of safe ML and interpreting classifiers respectively. 
\citet{Huang2017-lk} use `algorithmic teaching' (see~\cite{Balbach2009-jw}) as inspiration for helping human users learn a robot's true objectives. Algorithmic teaching involves having a model of a students learning algorithm, and then presenting training examples to allow the student to learn a target model. In this case the `student' is the human user, and the `teacher' is the robot that is trying to teach the human its own objective function by presenting a set of (optimal) training examples. Here we would consider the robot's training examples as assurances.

Another consideration is whether an explanation is meant to be descriptive or aimed at ensuring comprehension, as well as whether explanations need to be on a macro or micro scale relative for parts of the Bayesian network (similar to globally/locally interpretable learned models \cite{Ruping2006-xj}). 
\citet{Lacave2002-cu} address the AIA reduction of complexity from the perspective of explaining probabilistic inference in Bayesian networks---specifically, \emph{how} and \emph{why} a Bayesian network reaches a conclusion given some imputed evidence. 
They present three properties of explanation: 1) content (what to explain), 2) communication (how to explain), and 3) adaptation (how to adapt based on who the user is). Several key points for designing assurances arise from considering the differences between explaining evidence (i.e. data), the model (i.e. the Bayesian network itself), or the reasoning (i.e. the inference process).

\citet{Aitken2016-cv} propose a metric called `machine self-confidence' for providing users with better insight into autonomous decision making under uncertainty. This insight comes from breaking down the complex influences of uncertainty on the decision making process and presenting them to the user in a simple way. Self-confidence is defined as the machine's own perception of its ability to carry out tasks in light of uncertainties in its knowledge of the world, its own self/states, and its reasoning process and execution abilities. In this sense, self-confidence is an AIA's metacognitive assessment of its own behavior and `competency boundaries'. A computational measure for POMDP-based autonomous planning is defined from five component assurances (which are fairly general and applicable to most other kinds of planners): 1) Model Validity, 2) Expected Outcome Assessment, 3) Solver Quality, 4) Interpretation of User Commands, and 5) Past Performance. 
The key idea behind this set of measures is to assess where and when approximations required for planning under uncertainty are expected to break down. Model validity attempts to quantify the validity of a model within the current situation. The expected outcome assessment uses the distribution over rewards to indicate how beneficial or detrimental the outcome is likely to be. Solver quality quantifies how a specific POMDP solver is likely to perform in the given problem setting (i.e. how close to optimal the solution policy and approximate solution policy can get). The interpretation of commands component is meant to quantify how well the objective has been interpreted (i.e. how sure is the AIA that it correctly interpreted mission specifications into relevant tasks and suitable goals). Finally, past performance is meant to add in empirical experience from past missions, in order to make up for theoretical oversights and account for learning-based processes.

\citet{Aitken2016-cv} proposes that self-confidence values could, for instance, be reported as a single value between $-1$ (complete lack of confidence in achieving mission objectives) and $1$ (complete confidence in achieving mission objectives); a self-confidence value of $0$ reflects total uncertainty. Each of the component assurances could be useful on its own, but the composite `sum' of the factors is meant to distill the information from the five different areas, so that a (possibly novice) user can quickly and easily evaluate the ability of the AIA to perform in a given situation. Currently, only two of the five metrics (Expected Outcome Assessment, and Solver Quality) have been developed quantitatively, but there is continuing work on the other metrics and they plan to perform human experiments to evaluate the usefulness of the self-confidence metrics for AIAs. Other approaches for computing and communicating AIA self-confidence have also been proposed for more specific applications \cite{Hutchins2015-if, Kaipa2015-hy, Zagorecki2015-qy, Kuter2015-qh}.

%% file: visualization_dr.tex
\subsection{Information Visualization} \label{sec:vis_dr}
We define `information visualization' as the act of displaying artifacts generated by AIA models or processes for the task at hand in such a way as to communicate to one of the trust dimensions of a human user. Specifically we consider the `competence' and `predictability' of the AIA, as well as the `situational normality' of the task at hand. This can overlap but is not necessarily the same as generating self-assessments, which are introspective and \emph{process based} assurances (i.e. which are descriptive and reflective of the AIA's capabilities); rather, information visualization tends to more broadly include or revolve around \emph{outcome based} assurances (i.e. which focus on results or expected results of applying the AIA's capabilities).  

\subsubsection{Common Approaches:}
\citet{Liu2017-xw} review several of the current methods that exist for visualizing machine learning models. They identified three main purposes for which visualizations are useful in this context: 1) understanding (why models behave the way they do on certain problems), 2) diagnosis (failures, or unexpected behavior on certain tasks), and 3) refinement (ability to improve performance on tasks). 
We consider two common methods that assist in these processes: dimensionality reduction and visualization of uncertainty.

\begin{figure}[htpb]
    \centering
    \includegraphics[width=0.7\linewidth]{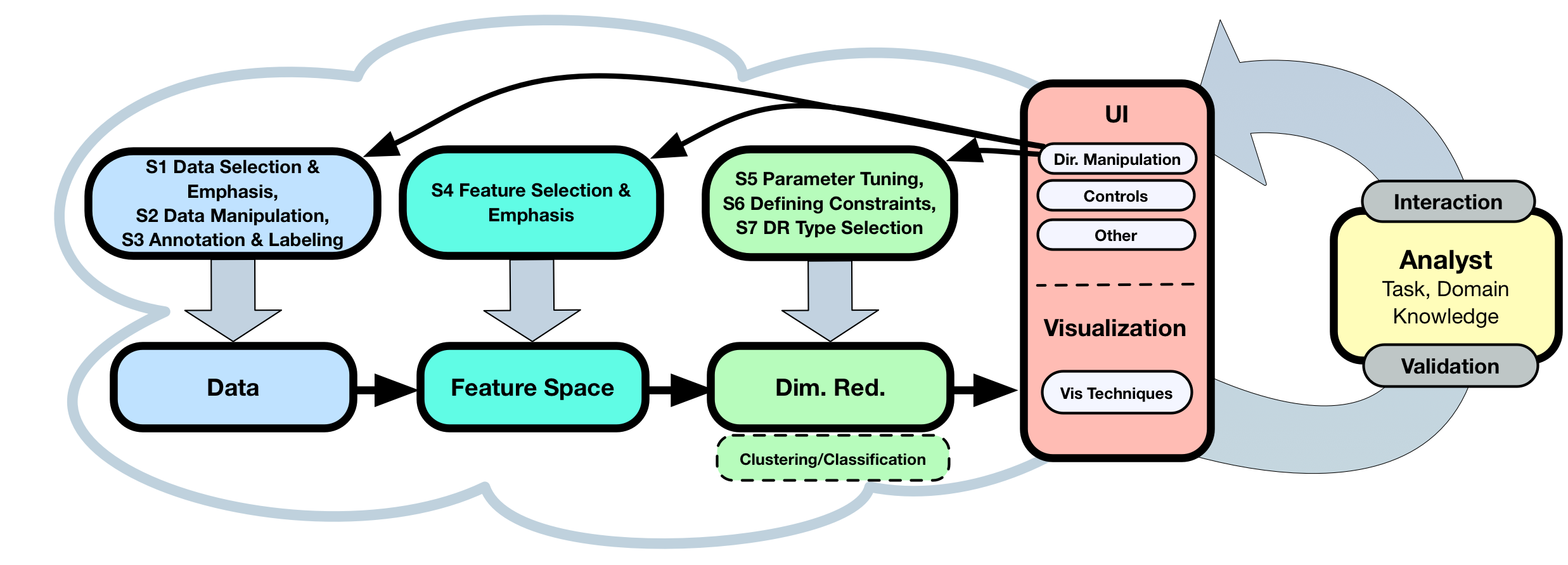}
    \caption{Human-in-the-loop process model ~\cite{Sacha2017-hf}~\textcopyright 2017 IEEE. Reprinted, with permission, from Sacha, Dominik, et al. ``Visual Interaction with Dimensionality Reduction: A Structured Literature Analysis.'' IEEE Transactions on Visualization and Computer Graphics, vol. 23, no. 1, Jan. 2017, pp. 241-50.}
    \label{fig:sacha_fig}
    \vspace{-0.2 in}
\end{figure}

\paragraph{Dimensionality Reduction:}
Dimensionality reduction (DR) is one of the key methods used in creating visualizations. \citet{Sacha2017-hf} identify seven different methods by which users interact with DR techniques. They use this to make the human-in-the-loop process model for interactive DR that is shown in Figure~\ref{fig:sacha_fig}. This interactive nature of their model helps users to better understand the information that they are viewing.

\citet{Venna2007-yj} discusses DR for ML and reviews many linear and non-linear projection methods. \citet{Vellido2012-nm} also discusses the importance of DR for making ML models interpretable. As one example, \citet{Chipman2005-om} applied this idea by constraining principle component analysis (PCA) in an attempt to make the resulting linear combinations of variables more interpretable (more homogeneous, or more sparse).

At times a simple visualization is the most efficient way to communicate the results of decision making or planning. For example: \citet{Chadalavada2015-wx} enable a robot to project its path onto the ground so users can see.

\paragraph{Treatment of Uncertainty:}
In the previous section we have already visited the importance of an AIA being able to quantify its uncertainty. Visualization researchers are concerned with how to \emph{convey} that uncertainty to human users (and quantify uncertainty inherent in making visualizations). \citet{Sacha2016-tu} discuss how the propagation of uncertainty through visual analytics systems can affect the trust of human users (see also \cite{Correa2009-hi}).
One excellent example of this is the work by \citet{Wu2012-qi}, who create a tool to visualize the flow and propagation of uncertainty in the visualization process. In this way users can understand where uncertainty enters the data visualization process.

\begin{figure}[htpb]
    \centering
    \includegraphics[width=0.6\linewidth]{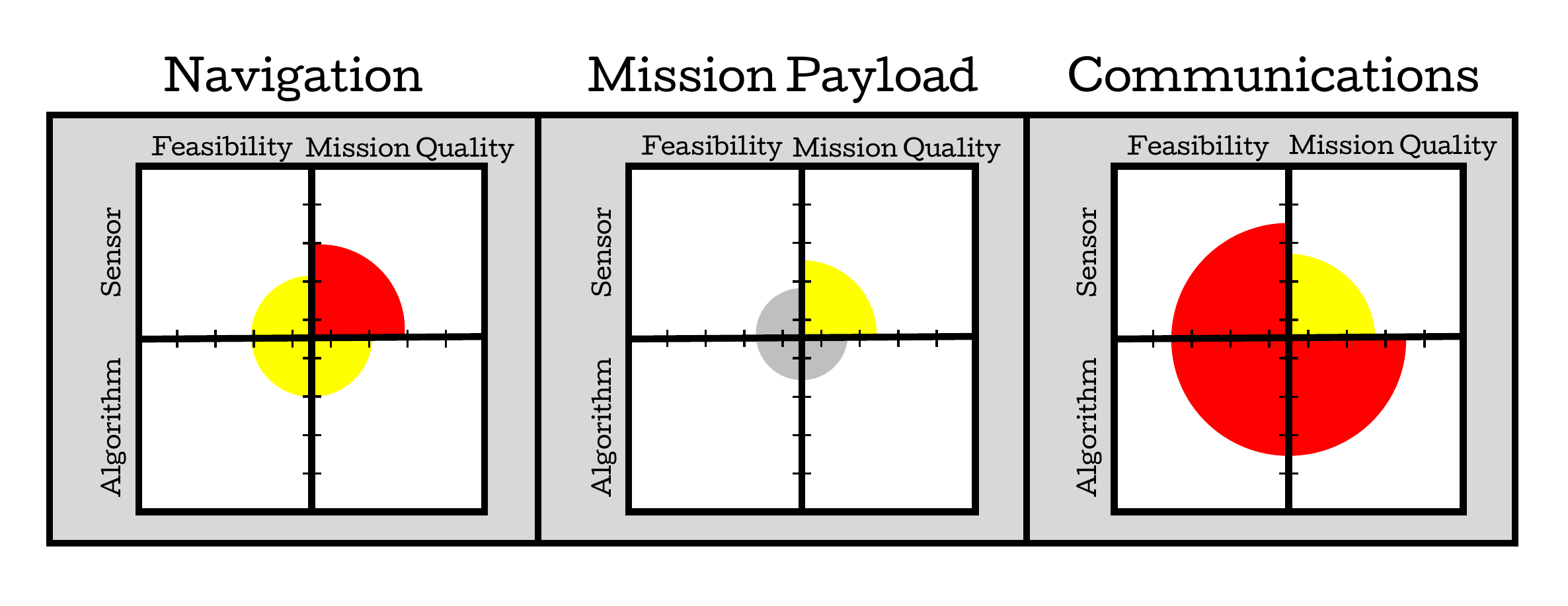}
    \caption{Proposed `trust annunciator panel'~\cite{Hutchins2015-if} Hutchins, Andrew R., et al., Proceedings of the Human Factors and Ergonomics Society Annual Meeting, ``Representing Autonomous Systems' Self-Confidence through Competency Boundaries'' Vol. 59, Issue 1, pp. 279-83. copyright \textcopyright 2015 by SAGE Journals, Reprinted by Permission of SAGE Publications, Inc.}
    \label{fig:hutchins_fig}
    \vspace{-0.2 in}
\end{figure}

The relationship between systemic uncertainties and their effects on system performance can be very complex. \citet{Hutchins2015-if} address this by using expert knowledge, and a `trust annunciator panel' (TAP) that has several `uncertainty level indicators' in order to display how uncertainties in sensors will effect the output quality, and the mission impact; and the same for the planning algorithm (see Figure~\ref{fig:hutchins_fig}).

\subsubsection{Grounding Example:}
In the case of the `VIP Escort' problem (described in Section~\ref{sec:mot_example}), information visualization might be used as an assurance in the following way, starting with the assumptions that:

\begin{itemize}
    \item The UGV has just begun an attempt to escape the road-network
    \item The user has access to an interface like that proposed in \cite{Hutchins2015-if}
\end{itemize}

During the attempt the user is able to see how the sensor uncertainty might possibly effect the outcome of the mission. In this case, the user is assured that the sensors will have little negative impact on the outcome of the mission given the current weather conditions.

\paragraph{\textbf{Discussion of Example:}} Here we see how a visualization is able to assist the user in correlating the effects between sensor uncertainty and mission outcome. This is not a simple relationship for operators (especially untrained) to learn on their own; even if they were able to learn the time required to do so can be very detrimental.

%% file: user_assessment.tex
\subsection{User Assessment} \label{sec:user_assessment}
In this section we address assurances that are based solely on user assessment; in other words the AIA expends little or no computational effort to `digest data' to turn into assurance information for the user, and instead relies the user's own cognitive abilities to draw assurances from their observations. Such assurances are clearly not integral to the function of the AIA, as they might, for example, be realized by incorporating simple displays, print statements, or other `raw data' indicators into a basic user interface. 
This category is in contrast to Section~\ref{sec:vis_dr}, where the AIA is designed to process data to assist the user in understanding how to trust the AIA appropriately. This approach is enticing for many system designers given how easy it is to implement at any stage of AIA design (even as an after-thought). However the effectiveness of this approach rests on several, strong, assumptions:

\begin{itemize}
    \item The user can form a `good enough' mental model of the AIA on their own to inform appropriate TRBs;
    \item Different users have `similar enough' capabilities and experiences to draw appropriate inferences on their own;
    \item There are no other compelling sources of information that will confound the assurance;
    \item Common cognitive biases won't interfere with the long-term operation/supervision of the system (e.g. recency, framing or anchoring effects that skew user's perception of non-linear changes in performance variables like power/fuel consumption). 
\end{itemize}

The weight of each of these assumptions relies heavily on the task to be performed, and the characteristics of the typical users. For example, in situations with highly trained personnel (i.e. military, or manufacturing facility) all users will have similar level of capability; thus `user assessment' is a viable and effective solution. 
In other scenarios with more diverse users and operating environments these assumptions begin to break down (i.e. mass market consumer products).

\subsubsection{Common Approaches:}
As suggested in the section's name users can form assurances by any method of perception. The most commonly investigated approaches are: simple, visual, `display of raw data'; and `by inspection' performance-based assessment. 

\paragraph{Display of Raw Data:}
Assurances associated with displaying AIA performance variables sound banal (e.g. flow rate for an automated pump \cite{Muir1996-gt}), but they actually make use of a nuanced point: the displayed performance value actually serves to inform the user's own mental model of the trustworthiness of an AIA capability. That is, the user's trust in the AIA's capability does not change only in response to the instantaneous `goodness/badness' of the AIA's performance, but accounts for the past history of the AIA's performance as well as any observed discrepancies between the AIA's expected behavior and its actual behavior. The user's trust dimensions (`competence', 'predictability', etc) are then affected by their perception of trustworthiness according to the combined model and data delivered by the display. This approach (also noted and discussed by \cite{Wickens1999-la,Sheridan1984-kx,Hutchins2015-if}) is effective, but relies heavily on the implicit assumption that the user will create a `good enough statistical model' of the AIA's behavior from data presented by the AIA. With this in mind, one might train a user to recognize signs of failure/success in different interactions with an AIA as assurances \cite{Freedy2007-sg,Desai2012-rc,Salem2015-md}. The main drawback of this idea is that it still relies on users' ability to construct `good enough' mental models of AIA behavior and characteristics from noisy observations to avoid misinterpreting AIA behaviors. It can also require intensive and costly special effort for non-expert or non-specialist users. A more ideal approach in such cases would be to design explicit assurances that help users construct correct/consistent mental process models of AIA behavior and thus reduce the risk of misinterpretation.

\paragraph{Performance-based:}
Users can also be assured by directly assessing the performance of an AIA on their own without any additional aiding or prompting.  Put simply: \emph{making stuff that (obviously) doesn't break improves trust}. \citet{Riley1996-qm} investigated how reliability and workload affected the participant's likelihood of trusting in automation. Two simulated environments were created to this end. First was for participants to use/not use an automated aid (with variable reliability) to classify characters while also performing a distraction task. Interestingly, they found that pilots (those with extensive experience working with automated systems) had a bias to use more automation, but reacted similarly to students in the face of dynamic reliability changes.

In a similar vein \citet{Desai2012-rc} investigated the effects of robot reliability on the trust of human operators. In this case, a human participant needed to work with an autonomous robot to search for victims in a building, while avoiding obstacles. The operator had the ability to switch the robot from manual (teleoperated) mode, to semi-autonomous, or autonomous mode depending on how they thought they could trust the system to perform. During this experiment the reliability of the robot was changed in order to observe the effects on the operator's reliance to the robot. Trust was measured by the amount of time the robot spent in different levels of autonomy (i.e. manual vs. autonomous), and it was found that trust changed based on the levels of reliability of the robot. \citet{Yu2018-qw} also had similar findings in their study of operators utilizing an `automatic quality monitor'.

\subsubsection{Grounding Example:}
In the case of the `VIP Escort' problem (described in Section~\ref{sec:mot_example}), user assessment might be used as an assurance in the following way, starting with the assumptions that:

\begin{itemize}
    \item The UGV has just begun an attempt to escape the road-network
    \item The user can observe the location of the UGV on the road network
    \item The user has access to the speedometer of the UGV
    \item The user has been trained and understands how the UGV functions
\end{itemize}

As the user monitors the UGVs progress they notice that, on a particular stretch of road, the speedometer reading seems very high, and the UGV stops moving. They recall from training that in situations where the speedometer shows a high speed and the UGV isn't moving it is likely that the UGV is spinning out or high-centered. They are able to diagnose the failure and dispatch the appropriate assistance.

\paragraph{\textbf{Discussion of Example:}} In this case the user was able to diagnose a problem based on the UGV not moving and the speedometer being high. They were able to do so because they were familiar with the system and were trained to be able to recognize this kind of situation. In future interaction the user \emph{might} associate the failure to certain characteristics of the road, or other properties of the task\ldots \emph{or} just feel like the UGV isn't very competent.

%% file: future_work.tex
The formal design of algorithmic assurances is still an emerging field. Consequently, there are many opportunities for further research along different lines. This section outlines some possible promising directions for future work.

\vspace{-0.15 in}

\subsection{Properties of Assurances} \label{sec:assurance_props_future}
 Figure~\ref{fig:refined_trust} gives some hints about how designers might be able to fully characterize the properties of AIA assurances. In this survey we investigate, in some detail, `Level of Integration'. However all of the other grayed-out boxes in Figure~\ref{fig:refined_trust} have open questions that should still be investigated. 

\input{ass_st.tex}

\input{ass_cc.tex}

\input{ass_ei.tex}

\input{ass_tt.tex}

\subsection{Trust and Distrust}
The treatment of assurances in this survey is based, in part, on a model of interpersonal trust. For completeness it will be important to further investigate \textit{distrust}, as reviewed and discussed by \citet{Lewicki1998-ox}, and formalized in \citet{McKnight2001-gz}. Low trust is not the same as distrust, and low distrust is not the same as trust. \citet{McKnight2001-gz} suggest that `the emotional intensity of distrust distinguishes it from trust', and they explain that distrust comes from emotions like wariness, caution, and fear -- whereas trust stems from emotions like hope, safety, and confidence. Trust and distrust are orthogonal elements that define a person's TRB towards a trustee. Since distrust was not considered here, it is not clear to what extent the human-AIA trust model remains effective in the presence of user wariness, caution, or fear. Questions for future work include: to what extent can behaviors driven by distrust be isolated from those originating from trust? How can those behaviors be detected to begin with? And in what circumstances is the extra effort necessary? 

\subsection{Human Limitations}
Dealing with human users requires consideration of their cognitive limitations. For instance cognitive biases known as `framing effects' (reacting to the same choice in different ways depending on how it is presented) will be important to consider for designing usable AIAs that must make decisions under uncertainty ~\cite{Freedy2007-sg,Riley1996-qm}. The existence of framing effects are not surprising to those familiar with cognitive science, but they will likely be unanticipated phenomena to many AIA system designers. Other related cognitive biases and limitations such as `recency effects' (being biased in making choices based on recent experience), `focusing effects' (being biased in choice selection based on a single aspect of a correlated event), or `normalcy biases' (failure to consider situations which have never occurred before) are also important to consider. 

Besides cognitive biases, humans are also limited in their ability to understand certain kinds of information. Communities that investigate how probabilistic and statistical explanations can be presented to humans will have many insights that are relevant for AIA designers and assurance design \cite{Rouse1986-dz,Wallace2001-fm,Kuhn1997-qc,Lomas2012-ie,Swartout1983-ko}. But it is not immediately clear what methods are most appropriate for application in assurance design, or how they might be applied. For instance, can the AIA detect when cognitive limitations are effecting TRBs? What other user limitations need to be characterized? 

\input{perception_mediums.tex}
\input{obs_effects.tex}

%% file: ass_st.tex
\emph{Source-Target Classification:}
It would be especially convenient for designers of AIAs to be able to refer to assurances by way of their source AIA capability (see Figure \ref{fig:AIcapabilities}) and their target user trust dimension (see Figure \ref{fig:Assurance_classes}). 
For example, human-like gestures could be considered a `motion-predictability' assurance. For instance, the reader may be able, in retrospect, to identify work from \cite{Dragan2013-wd} as a `motion-predictability' assurance, while the work by \citet{Wang2016-id} could be considered to describe `perception-competence' and `planning-predictability' assurances (among others). Meanwhile, \citet{Aitken2016-fb} considered a large set of assurances that span several source capabilities, and target trust dimensions. 
However, it is not always easy to clearly separate AIA capabilities or trust dimensions due to the inherent cross-over.  

Still, such classifications are useful because different classes of algorithms will likely present themselves as useful in applications for which assurances must target `predictability' dimensions of trust, for example, as opposed to `situational normality'. 
Given the inherent difficulty of precisely modeling and measuring trust, it is not immediately clear how such mappings can be precisely delineated.  
Future research might begin by looking for missing correspondences or correlations in the literature for notional capability source-trust target pairs. For example, have satisfactory assurances been developed for the `learning-situational-normality' source-target pair? Or, to what extent (if any) can assurance $x$ for `perception-competence' also be applied to `learning-competence'? Finally, are there certain classes of algorithms that are suited for communicating to the `predictability' dimension of trust, and can they be adapted from one AIA capability to another?

%% file: ass_cc.tex
\emph{Component and Composite:}
A component assurance is an assurance that describes a single AIA capability (or one aspect of an AIA model/process for a capability), and targets one or more trust dimension targets.  Component assurances are the most well researched in the existing literature. A component assurance might include displaying the confidence of a classification prediction, or visualizing a model. A composite assurance is the combination of more than one component assurance into a single assurance. 
A notable example is the machine self-confidence work by \citet{Aitken2016-cv} which notionally combines five component assurances into a single composite assurance interpretable to non-experts. 

\begin{figure}[!htbp]
    \centering
    \includegraphics[width=0.7\textwidth]{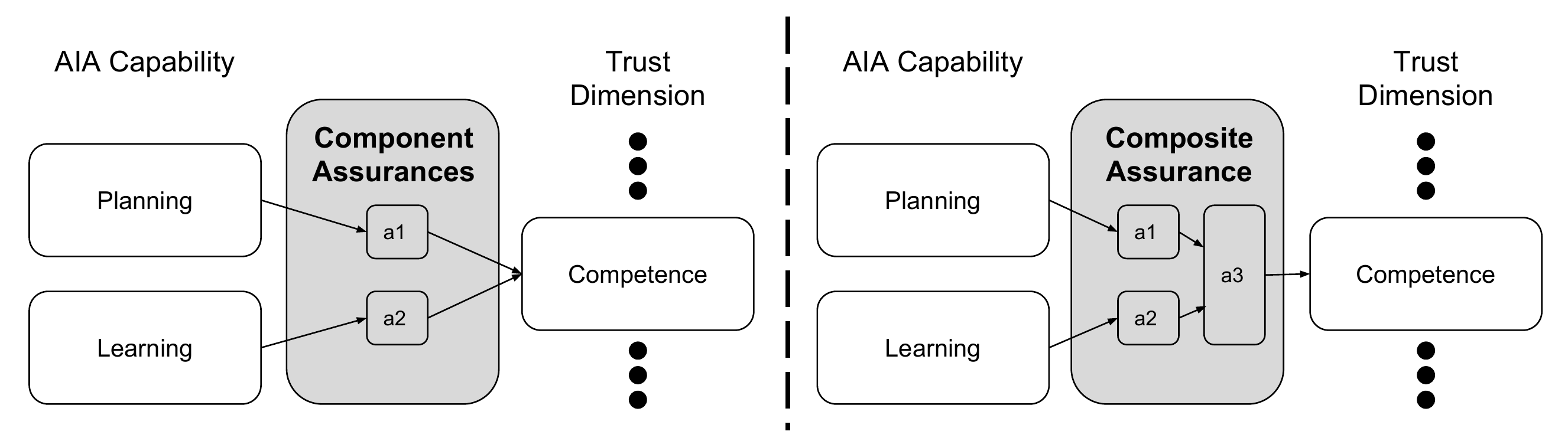}
    \caption{
    Component vs. composite assurances: the combination of multiple component assurances into a single assurance is a composite assurance. On the left are two component assurances $a1$ and $a2$; on the right only $a3$ affects the trust dimension.}
    \label{fig:assurance_mapping}
    \vspace{-0.1 in}
\end{figure}

Figure \ref{fig:assurance_mapping} illustrates the concepts of component and composite assurances. Some open questions here are: how can component assurances generally be combined to create a composite one? Also, to what extent can component and composite assurances be used in concert to provide assurances for users of differing expertise? And, are the effects of a composite assurance equal to the sum of its components?

%% file: ass_ei.tex
\emph{Explicit and Implicit Assurances:}
This work has only considered \emph{designed algorithmic assurances}. However, users will always form some kind of trust relationship to an AIA, even if deliberately designed assurances are not available. In the absence of designed assurances, user trust is informed by \emph{implicit} undesigned assurances. These can be thought of as artifacts or side-effects of other design decisions not meant to directly influence user trust.

Why is it important to consider implicit assurances? There is always a danger that users attend to the `wrong' assurances, i.e. AIA features that are not meant to be interpreted as assurances but are nevertheless easily perceived as such (possibly more so than intended explicit assurances). For example, a designer may create a planning-predictability assurance for an autonomous wheeled mobile robot, which could be rendered ineffective by an implicit assurance given by the appearance of that robot, e.g. the user may trust it less if the robot has old tires or has a large tool attached to its front end which makes it `look unsafe'. 

It remains an open question as to how designers can identify and mitigate the impact of implicit assurances, especially so that they do not confound the intended effects of explicit assurances. 
User studies will undoubtedly be helpful in obtaining feedback about which AIA characteristics most affect user trust, e.g. if explicit assurances are being perceived, and if there are implicit assurances whose effects overwhelm those of explicitly designed assurances. With such feedback, designers would have a realistic idea about whether their explicitly designed assurances are having the desired effect on user TRBs. 
However, the design and analysis of this issue remains open for further study, and is likely to have many application-specific dependencies (though, in the spirit of this paper, cross-domain comparisons would also likely prove valuable).

%% file: ass_tt.tex
\emph{Tutoring vs Telling:}
Assurances investigated to date are largely designed for one-way `telling' of information, i.e. that they do not consider and adapt to the experience or other traits of different users. The ability to adapt to different users, and tutor them to appropriately trust AIAs will become more critical as time passes, due to the diversity of user bases for advanced AIAs and time that users will spend interacting with them on complex tasks. A tutoring assurance might, for example, be a planned dynamic sequence of assurances that would change in time to adapt to the user's needs via two way user-AIA communication. This might include modification of assurances to help a user avoid boredom or fatigue in long-duration applications requiring user supervision, or to use the system differently in varying circumstances. It is not surprising that, to our knowledge, no research has been done with respect to tutoring a user in a trust relationship. This is a complex problem that requires understanding how different users learn and identifying potential strategies for eliciting appropriate TRBs from them. However, many interesting avenues for pursuing these ideas may come from the work on educational tutoring systems \cite{Wenger2014-ld} and algorithmic teaching  \cite{Balbach2009-jw}.

%% file: perception_mediums.tex
\subsection{Expression and Perception of Assurances} \label{sec:express_assurances}
Although specific algorithms can be used to build the contents of assurances, it is also critical to consider the actual communication of assurances. The expression (and subsequent perception) of an assurance involves considering mediums, methods, and efficacy. The medium of an assurance includes the form in which it expressed, e.g. visually, audibly, or otherwise. 
The method of expression includes for example using a plot, or a natural language phrase (which could be text-based or speech-based, depending on the medium).  Finally, the factors influencing the efficacy of the assurance must also be considered (e.g. consider using an audible assurance in a noisy environment). Humans generally utilize different methods/mediums when communicating assurances to each other to maintain efficacy when potential `losses in transfer' might occur. 
However, arguably the greatest challenge in using different mediums and methods is not in their implementation, but in designing the ability to recognize and decide when they should be applied. Some interesting questions are: In what circumstances are different methods most useful? And the same for mediums? How can different methods/mediums be selected in order to maximize assurance efficacy while also taking into account that using all possible combinations will \emph{not} help the user? How, and to what extent, can AIAs assess the efficacy of an assurance before, during, or after operation?

%% file: obs_effects.tex
\subsection{Observing Effects of Assurances} \label{sec:measuring_effects}
    Since assurances are meant to influence TRBs, it is important to quantify these effects so that:  1) the AIA system designer can understand how effective the assurances actually are; and 2) the AIA can evaluate the efficacy of its assurances and adapt them as needed.
    To our knowledge, there has not been any work that enables an AIA to observe user responses to assurances and then adapt behaviors appropriately (at least not in the trust cycle setting). 

    There are two known approaches to measuring the effects of assurances: gathering self-reported changes~\cite{Mcknight2011-gv,Muir1996-gt,Wickens1999-la,Salem2015-md,Kaniarasu2013-ho}, and measuring changes in TRBs~\cite{Freedy2007-sg,Desai2012-rc,Salem2015-md,Wu2016-ei,Bainbridge2011-pl}. Measuring changes in TRBs is the more objective approach generally speaking, but the choice between one method and the other depends on the application. Still, more investigation is needed to identify the \emph{principles} behind measuring the effects of assurances. Some interesting, yet unanswered, questions include: are there some TRB measurement strategies that fare better than others for particular kinds of applications or assurances? In what ways, if any, do these methods need to be adapted to suit different kinds of users? Is it possible to show that there are in fact causal relations from specific assurances to specific TRBs?

%% file: conclusions.tex
The issues of user trust in AIAs and appropriate deployment/use of AIAs have become very prominent.  Assurances are the method by which AIAs can influence humans to trust and (more importantly) \emph{use} them appropriately. We have presented here a definition, case for, and survey of algorithmic assurances in the context of human-AIA trust relationships. A formal treatment of this topic is necessary because the ecosystem of AIAs is evolving more rapidly than ever before; consequently, previous informal approaches to designing algorithmic assurances are insufficient. 

This survey was performed, to some extent, from a standpoint of designing intelligent unmanned vehicle systems that must work in concert with a human supervisor. However, the theoretical framework and categorization of assurances is meant to be generally applicable to a broad range of AIAs. A major motivation for this survey was the observation that there are many researchers in different but related domains such as human factors, robotics, machine learning, artificial intelligence, and others who are (unknowingly) working along different parts of the same human-AIA assurance spectrum. It is important for members of each community to recognize this, so that research efforts can be  methodically organized to answer related open questions in this important area. Assurances have historically been ignored from a practical standpoint, and are the least understood component of human-AIA trust relationships. There have been many researchers who have recognized the concepts behind assurances, but no detailed definitions have been given until now.

There are three main contributions from this work: 1) we have drawn from multiple bodies of research in order to fill in the missing details for the human-AIA trust cycle (Fig.~\ref{fig:SimpleTrust_one_way}) and to formally define assurances within this cycle; 2) we present a classification of assurances in Sec.~\ref{sec:assurances}; 3) we identify an `assurance integration continuum' shown in Fig.~\ref{fig:assurance_continuum}. On that continuum seven different classes of algorithms were identified. Practitioners can use these classes to select and design assurances for AIAs. Given the material provided herein, those who design assurances should have the tools required to approach design and future research from a solid theoretical foundation.

A final important and sobering takeaway is that there is not a single `silver bullet' algorithmic assurance that will perform the best in all situations. 
Given enough time, it is quite possible that highly specialized assurances could be designed for many situations. Even so, we warn that, for the research and design of assurances to be sustainable in the current environment of fast-paced development of new technology, it is important to consider approaches that are as principally grounded as possible, in order to be more easily used with yet-to-be-invented methods for implementing various AIA capabilities. We have identified many future opportunities for research on AIA assurance design and their influence on human trust, and hope researchers will begin looking outside of their own disciplines to discover, design and formally test new tools and ideas for assurance design and implementation. The framework presented here should unify research efforts by providing a common taxonomy in relation to human-AIA trust relationships. We believe it will help researchers see the field from a larger perspective, classify the type of research they are performing, and consider the greater implications of their work. The field of algorithmic assurances has an abundance of avenues for new and challenging research, and we encourage researchers to pursue them.